\documentclass[structabstract]{aa}
\pdfoutput=1
\usepackage[pdftex]{graphicx}
\usepackage{natbib,amssymb,amsmath,url}
\usepackage{hyperref}
\usepackage{color}

\newcommand{\plotwd}{9.cm}
\newcommand{\plotwdtwo}{18.cm}

\begin{document}
\title{Linking X-ray AGN with dark matter halos: a model compatible with AGN luminosity function and large-scale clustering properties}
\titlerunning{Linking X-ray AGN with dark matter halos}
%\authorrunning{G. H\"utsi et al.}
\author{Gert H\"utsi \inst{1,2}, Marat Gilfanov \inst{1,3}, Rashid Sunyaev \inst{1,3}}
\institute{Max-Planck-Institut f\"ur Astrophysik, Karl-Schwarzschild-Str. 1, 85741 Garching, Germany \\ \email{gert@mpa-garching.mpg.de} \and Tartu Observatory, T\~oravere 61602, Estonia \and Space Research Institute of Russian Academy of Sciences, Profsoyuznaya 84/32, 117997 Moscow, Russia}
\date{Received / Accepted}

\abstract
%{We investigate the luminosity function and large-scale clustering bias of the X-ray selected AGN.}
{}
{Our goal is to find a minimalistic model that describes the luminosity function and large-scale clustering bias of X-ray-selected AGN in the general framework of the concordance  $\Lambda$CDM model.
}
{We assume that a simple population-averaged scaling relation between the AGN X-ray luminosity  $L_{\rm X}$ and the host dark matter halo mass $M_{\rm h}$ exists. With such a relation, the  AGN X-ray luminosity function can be computed from the halo mass function. Using the concordance $\Lambda$CDM halo mass function for the latter, we obtain  the $M_{\rm h}-L_{\rm X}$ relation required to match the redshift-dependent AGN X-ray luminosity function known from X-ray observations. 
}
{We find that with a simple power-law-scaling $M_{\rm h}\propto L_{\rm X}^{\Gamma(z)}$, our model can successfully reproduce the observed X-ray luminosity function.  
Furthermore, we automatically obtain predictions for the large-scale AGN clustering amplitudes and their dependence on the luminosity and redshift, which seem to be compatible with AGN clustering measurements. Our model also includes the redshift-dependent AGN duty cycle, which peaks at the redshift $z\simeq 1$, and its peak value is consistent with unity, suggesting that on average  there is no more than one AGN per dark matter halo. For a typical X-ray-selected AGN at $z\sim 1$, our best-fit $M_{\rm h}-L_{\rm X}$ scaling implies low Eddington ratio $L_{\rm X}/L_{\rm Edd}\sim 10^{-4}-10^{-3}$ (2--10 keV band, no bolometric correction applied) and correspondingly high mass-growth e-folding times, suggesting  that typical X-ray AGN are dominantly fueled via relatively inefficient `hot-halo' accretion mode.}
{}
\maketitle

\section{Introduction}
That every galaxy of significant size harbors a supermassive black hole (SMBH) (with masses $\sim 10^6-10^9\,M_\odot$) is probably one of the most remarkable discoveries of modern astrophysics. The growth of SMBHs, as manifested by the active galactic nuclei (AGN), has been observed over a broad range of electromagnetic energies, from radio to hard X-rays and gamma-rays. The observed AGN X-ray emission, believed to stem from the upscattering of softer accretion disk photons via inverse Compton mechanism by the hot electron corona, has proven to be the most effective way of selecting large samples of AGN over large cosmological volumes~\footnote{In comparison, the deepest optical spectroscopic surveys typically give a factor of $\sim 10$ times less AGN per deg$^{-2}$, and only utradeep optical variability studies are able to generate comparable AGN sky densities \citetext{e.g.,~\citealp{2005ARA&A..43..827B}}.}. This efficiency comes from the fact that to absorb X-rays one needs significant column densities of absorbing material, which is particularly true in the hard X-ray band, where $N_{\rm H}\sim10^{24}$ cm$^{-2}$ is required.

With XMM-Newton~\footnote{\url{http://xmm.esac.esa.int}} and Chandra~\footnote{\url{http://chandra.harvard.edu}} X-ray observatories, two of the most advanced X-ray instruments in existence, more than 20 deep extragalactic X-ray surveys have been performed over varying sky areas and limiting sensitivities \citep{2005ARA&A..43..827B}. The most noticeable among these are $\sim 2$ Ms Chandra Deep Field North \citep{2003AJ....126..539A} over $\simeq 448$ arcmin$^2$ sky area, $\sim 4$ Ms Chandra Deep Field South \citep{2011ApJS..195...10X} over $\simeq 465$ arcmin$^2$, and $\sim 3$ Ms XMM-Newton deep survey covering Chandra Deep Field South \citep{2011A&A...526L...9C}.

Despite the small sky areas covered by those deep surveys, a relatively large number of detected AGN has allowed reasonably good determination of the AGN luminosity function (LF) and its evolution over cosmic time \citep{2003ApJ...598..886U,2005A&A...441..417H,2010MNRAS.401.2531A}. Even though these narrow survey geometries are not very suitable for measuring the spatial clustering properties of AGN, somewhat wider and shallower surveys like X-Bo\"otes, XMM-LSS, AEGIS, and XMM-COSMOS have allowed measuring the two-point clustering statistics \citep{2005ApJS..161....1M,2006A&A...457..393G,2009ApJ...701.1484C,2009A&A...494...33G,2011ApJ...736...99A}, albeit with relatively large uncertainty on large scales. The large-scale clustering measurements are expected to improve significantly with the upcoming Spectrum-X-Gamma/eROSITA~\footnote{\url{http://www.mpe.mpg.de/erosita/}}\footnote{\url{http://hea.iki.rssi.ru/SRG/}} space mission \citep{2010SPIE.7732E..23P}, which is planned to cover entire sky down to the limiting sensitivity of $\sim 10^{-14}$ erg/s/cm$^2$ in the $0.5-2$ keV band \citep{2012arXiv1209.3114M,2012arXiv1212.2151K}. A good summary of the current status and future prospects for the X-ray-selected AGN clustering can be found in \citet{2012AdAst2012E..25C,2013arXiv1308.5976K}.

Even though the evolution of AGN over cosmic time is an interesting subject on its own, the remarkable discoveries of correlations between SMBH masses and host galaxy properties \citetext{see, e.g.,~\citealp{1995ARA&A..33..581K,2005SSRv..116..523F}, for reviews}, hinting at the underlying AGN-galaxy co-evolution mechanisms, have significantly broadened the importance and actuality of those topics in the astronomical community. Although controversial in terms of observational data \citetext{e.g.,~\citealp{2002ApJ...578...90F,2011Natur.469..377K}}, in currently favored cold dark matter (CDM) cosmologies, where the structure grows in a hierarchical fashion starting from the smallest scales (i.e., the bottom-up scenario), one may expect that SMBH masses $M_{\rm BH}$ should also correlate with the masses of the host dark matter halos $M_{\rm h}$. Indeed, this is what one commonly finds in semi-analytic AGN-galaxy co-evolution models, e.g.,~\citet{2012MNRAS.419.2797F}.  There is a significant scatter in the $M_{\rm BH}$-$M_{\rm h}$ relation, suggesting  that there are other important parameters beyond $M_{\rm h}$ that determine $M_{\rm BH}$, for example, the merger history of the dark matter halo. However, it is still reasonable to use $M_{\rm h}$ as a single proxy for $M_{\rm BH}$,  linking the latter to the quantities for which we have good theoretical predictions available, such as halo mass function (MF) and clustering bias. The black hole mass alone does not determine the AGN luminosity, since the particular process that fuels  the black hole is also important.  However, in some circumstances, for example in the `hot-halo' fueling model, the latter may lead to a  tight correlation between AGN luminosity and the host halo mass.

Future wide-field X-ray surveys like eROSITA promise to give us very precise AGN clustering bias measurements as a function of redshift and luminosity \citep{2013arXiv1305.0819K}. Since clustering bias provides direct information on how AGN populate dark matter halos, this also helps in determining the dominant X-ray AGN fueling mode: `hot-halo' \emph{vs} `cold gas' accretion. As an example, semi-analytic models of~\citet{2012MNRAS.419.2797F} suggest that the hot halo fueling mode has quite strong a dependence between the AGN X-ray luminosity and host halo mass, with brighter AGN populating preferably higher mass (and thus more strongly biased) halos. For the cold gas accretion, on the other hand, which is presumably responsible for most of the quasar activity, the host halo mass and AGN luminosity are only weakly dependent, with most of the activity occurring in somewhat lower mass halos.

Studies combining the AGN LF and their clustering strength have been carried out by several authors using mostly optically selected quasars (QSOs), such as \citet{2001ApJ...547...12M,2001ApJ...547...27H,2005ApJ...621...95W,2006ApJ...641...41L,2009MNRAS.395.1607W,2010MNRAS.404..399B,2010MNRAS.406.1959S,2010ApJ...718..231S,2013ApJ...762...70C}. The major differences between X-ray selected AGN and optically selected QSOs is that QSOs represent the brightest and the rarest end of the AGN population, while the X-ray-selected AGN, as detected in the deep X-ray surveys, are much more numerous and thus more typical examples of AGN. The other important difference is that despite QSOs being much rarer, their clustering strength is somewhat lower than the clustering amplitudes typically obtained for the X-ray selected AGN: QSO clustering is compatible with being practically independent of luminosity and having a similar strength as the dark matter halos with masses $\sim 10^{12} h^{-1}M_\odot$, while the clustering bias of X-ray selected AGN suggests that they populate group-sized halos with typical masses of $\sim 10^{13} h^{-1}M_\odot$. Furthermore, the QSO activity is compatible with being driven by mergers of lower mass dark matter halos with plenty of available cold gas or by the accretion disk instabilities.

There is evidence that AGN clustering strength might scale with X-ray luminosity \citep{2012ApJ...746....1K,2012MNRAS.tmp..101K}. The same trend is also seen in semianalytical galaxy-AGN co-evolution models of \citet{2012MNRAS.419.2797F}, where in the `hot-halo' AGN fueling mode they find  quite a tight correlation between dark matter halo mass (and thus clustering bias) and AGN luminosity. It should be mentioned, however, that other authors analyzing similar samples of X-ray detected AGN, such as \citet{2009ApJ...701.1484C} and \citet{2013MNRAS.430..661M}, find no evidence of such luminosity dependence of  AGN clustering. 
This is not surprising given the rather poor accuracy of clustering measurements in the X-ray band. Indeed, currently available samples of X-ray-selected AGN typically contain between a few hundred and a few thousand objects.

Motivated by these ideas, we assume that a relation exists between the dark matter halo mass and the AGN X-ray luminosity it harbors, and therefore, the X-ray luminosity function of  AGN  can be computed from the mass function of dark matter halos. In comparison to the QSO samples with typical sizes of $\sim 100,000$ objects, the currently available X-ray-selected samples are about two orders of magnitude smaller, which which suggests that these models should be kept from being too complex. Thus, we initially assume that the $M_{\rm h}-L_{\rm X}$ scaling is given by a simple deterministic relation.
The search for the functional form of this relation that is consistent with the concordance $\Lambda$CDM halo mass function and the observed  X-ray AGN LF is presented in Section~\ref{sec2}. It turns out that this way we indeed obtain an acceptable fit to the X-ray AGN LF and, as a bonus, automatically obtain a prediction about X-ray AGN clustering properties, which seems to agree with available observational data. This and other consequences of our simple model are discussed  in Section~\ref{sec3}. In Section~\ref{newsec} we investigate how an increased level of the model complexity influences our previous results. In particular, we extend our model by allowing a scatter in the $M_{\rm h}-L_{\rm X}$ scaling relation and luminosity dependence of the AGN duty cycle. We bring our conclusions in Section~\ref{sec4}.

Throughout this paper we assume a flat $\Lambda$CDM cosmology with $\Omega_{\rm m}=0.27$, $\Omega_{\rm b}=0.045$, $h=0.70$, and $\sigma_8=0.80$.

\section{Mapping halo MF to X-ray AGN LF}\label{sec2}

In the recent semianalytic galaxy-AGN co-evolution model of \citet{2012MNRAS.419.2797F}, the authors consider two main AGN fueling modes: `hot-halo' and `starburst' mode. Some of the results produced by these models, which were presented in \citet{2012MNRAS.tmp..101K}, are plotted in Fig.~\ref{fig5}. Here the dark-shaded stripe represents the `hot-halo' mode, while the light-shaded region corresponds to for the sum of `starburst' and `hot-halo' modes. The starburst mode, where the SMBH fueling is induced by mergers or disk instabilities and where there is a lot of cold gas available, is arguably the one responsible for the QSO activity. On the other hand, once SMBH gets incorporated into group- or cluster-sized dark matter halo, which is filled with hot low density gas, the AGN fueling takes on a different character. In this case, in addition to metallicity, the gas cooling rate is largely dictated by the gas density and temperature, which under the assumption of hydrostatic equilibrium are directly related to the host dark matter halo mass. The gas cooling must in practice be moderated by the feedback effects from the SMBH.
Thus, in the `hot halo' fueling mode the mass of the host halo plays an important role in determining the amount of available fuel and its accretion rate, leading to the existence of a rather tight relation between the host dark matter halo mass and AGN luminosity, as illustrated by  Fig.~\ref{fig5}.

Motivated by these considerations, we start by assuming that there is a following simple scaling relation between the halo mass and AGN X-ray luminosity
\begin{equation}\label{eq1}
M_{\rm h}=M_0\left(\frac{L_{\rm X}}{L_0}\right)^{\Gamma(z)},
\end{equation}
where we take $L_0=10^{41}$ erg/s. Having the above scaling relation, we can immediately write for the AGN X-ray LF:
\begin{equation}\label{eq2}
\frac{{\rm d}n}{{\rm d}L_{\rm X}}(L_{\rm X},z)=f_{\rm duty}(z)\frac{{\rm d}M_{\rm h}}{{\rm d}L_{\rm X}}(L_{\rm X},z)\frac{{\rm d}n}{{\rm d}M_{\rm h}}\left[M_{\rm h}(L_{\rm X},z),z\right]\,,
\end{equation}
where ${\rm d}n/{\rm d}M_{\rm h}$ is the concordance $\Lambda$CDM model halo MF and $f_{\rm duty}$ the `duty cycle', in our case defined as a fraction of halos that contain AGN in its active state. For the halo MF we use the analytic form given by \citet{1999MNRAS.308..119S} and the hard band ($2-10$ keV) measurements from \citet{2010MNRAS.401.2531A} for the X-ray AGN LF. The \citet{2010MNRAS.401.2531A} LF data points in nine redshift bins are shown in Fig.~\ref{fig3}. We assume, somewhat simplistically, that LF measurements in all luminosity bins are statistically independent with the errors following Gaussian distribution; i.e., the likelihood function gets factorized and has a simple analytic form. To sample the parametric likelihood function we used Markov chain Monte Carlo (MCMC) methods, in particular the Metropolis-Hastings algorithm \citep{Metropolis53,Hastings70}. For finding the maximum likelihood point in parameter space we used the downhill simplex method as described in~\citet{1992nrfa.book.....P}.

\begin{figure}
\centering
\includegraphics[width=\plotwd]{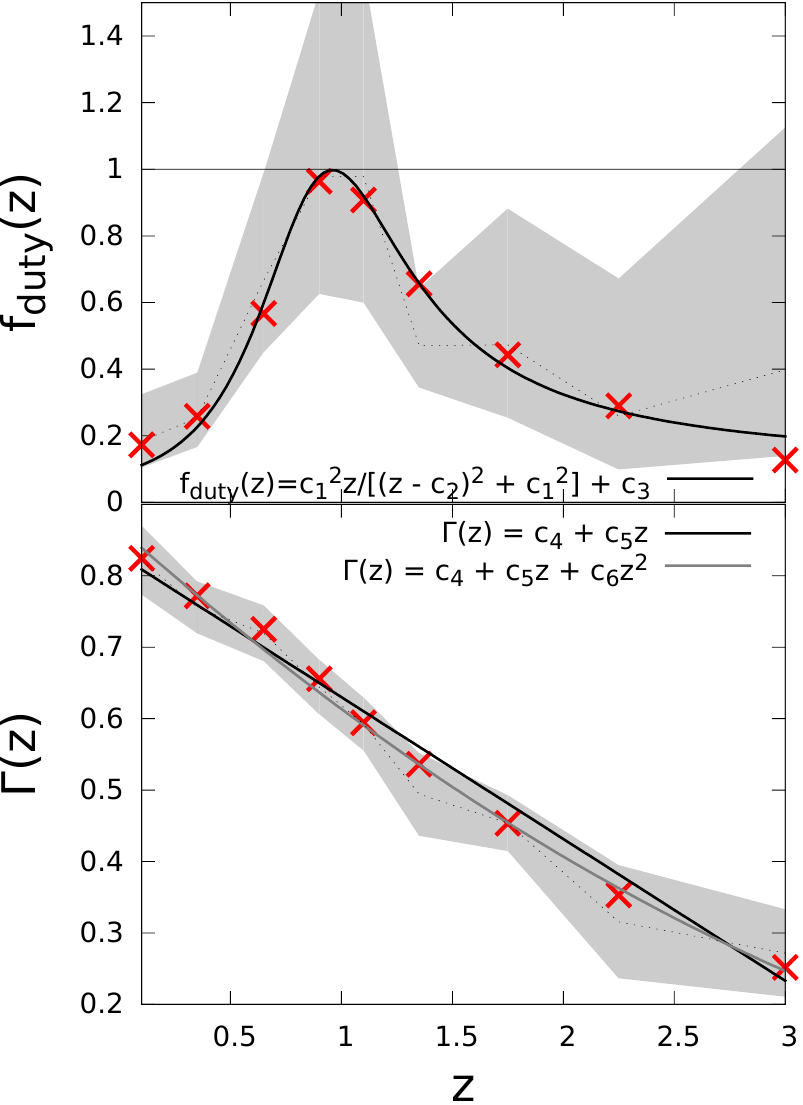}
\caption{Best fitting duty cycle values $f_{\rm duty}(z_i)$ (\emph{upper panel}) and power law indices $\Gamma(z_i)$ from Eq.~(\ref{eq1}) (\emph{lower panel}) in nine redshift bins, shown with red crosses. In the \emph{upper panel} the power law index is assumed to be in parametric form $\Gamma(z)=c_4+c_5z$ while the amplitude factors in each redshift bins are assumed to be free parameters, i.e., including $M_0$ the fitting problem has 12 free parameters in total. With solid black line we show the best fitting three parameter analytic function with the detailed form as specified in the legend. In the \emph{lower panel} $f_{\rm duty}(z)$ is fixed to the above analytic form while the power law indices in each redshift bin $\Gamma(z_i)$ are allowed to vary freely, i.e., this time we have 13 free parameters. Solid black and gray lines show linear and quadratic polynomial approximations to best fitting $\Gamma(z_i)$. The dotted lines and gray shaded regions represent best fit values along with $1\sigma$ intervals in case all $\Gamma(z_i)$ and $f_{\rm duty}(z_i)$ are all allowed to be free (19 free parameters).}
\label{fig1}
\end{figure}

\begin{figure*}
\centering
\includegraphics[width=\plotwdtwo]{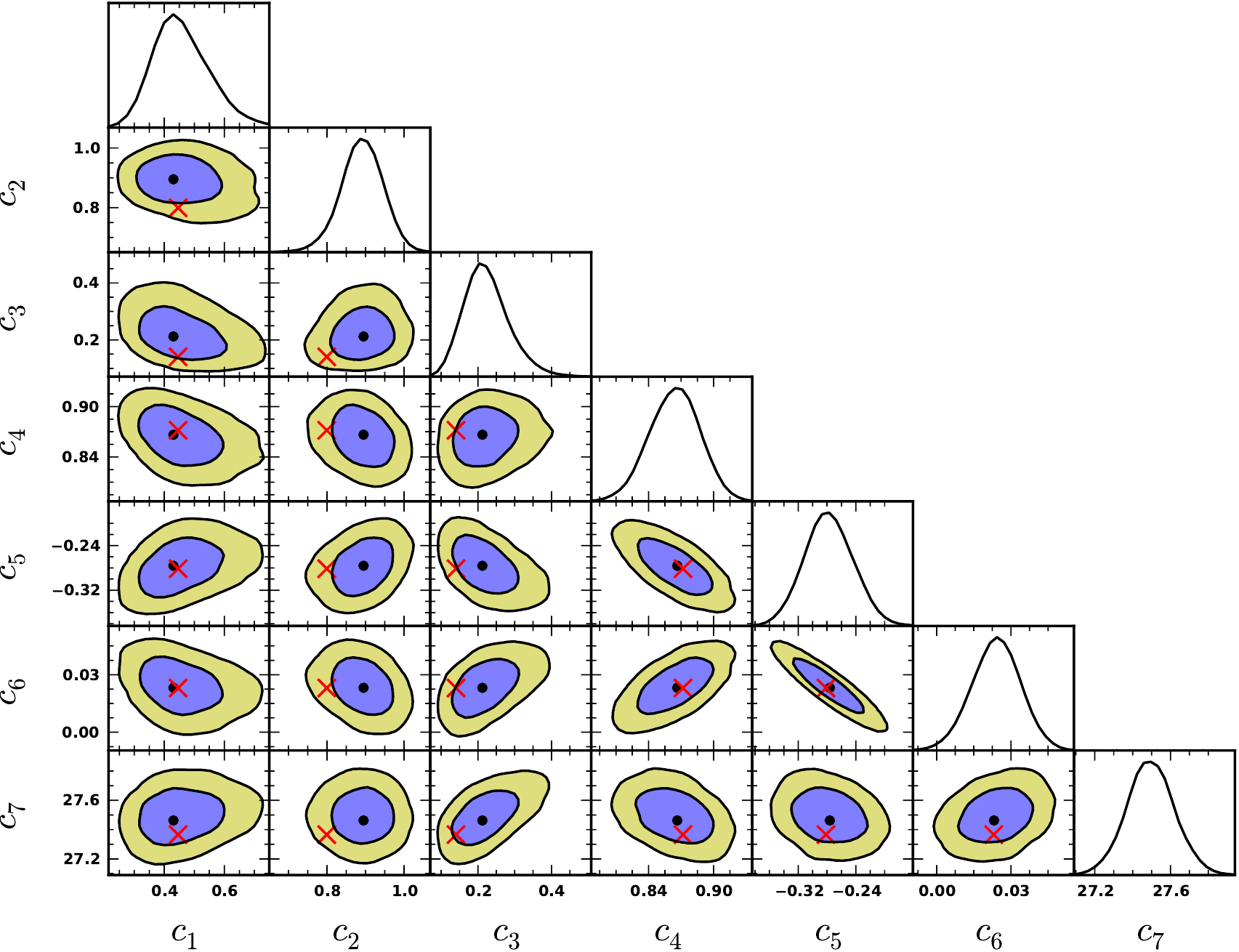}
\caption{Marginalized 2D error ellipses representing $1\sigma$ and $2\sigma$ confidence levels along with 1D marginalized probability distributions (upper panels in each of the columns) for all the free parameters of unconstrained Model II. The black dots and red crosses show the maximum likelihood points for unconstrained and constrained (i.e., $f_{\rm duty}\le 1$) Model II, respectively.}
\label{fig2}
\end{figure*}

\begin{table*}
\centering
\caption{Best fitting parameter values for Models I and II. Both constrained and unconstrained cases.}
\label{tab1}
\begin{tabular}{c|c|c|c|c|}
{\tiny $f_{\rm duty}(z)=\frac{zc_1^2}{(z-c_2)^2+c_1^2}+c_3$} & Model I ($c_6 \equiv 0$) & Model I ($c_6 \equiv 0$) & Model II & Model II\\
% & Model I ($c_6 \equiv 0$) & Model I ($c_6 \equiv 0$) & Model II & Model II\\
{\tiny $M_{\rm h}=M_0\left(\frac{L_{\rm X}}{L_0}\right)^{c_4+c_5z+c_6z^2}$} & + {\scriptsize $f_{\rm duty}\le 1$} &  & + {\scriptsize $f_{\rm duty}\le 1$} & \\
%{\scriptsize parameter} & + {\scriptsize $f_{\rm duty}\le 1$} &  & + {\scriptsize  $f_{\rm duty}\le 1$} & \\
\cline{2-5}
{\tiny $L_0=10^{41}$ erg/s} & {\scriptsize best fit}  & {\scriptsize best fit}  & {\scriptsize best fit}  & {\scriptsize best fit}\\
% & {\scriptsize best fit} & {\scriptsize best fit}  & {\scriptsize best fit}  & {\scriptsize best fit} \\
\hline
{\large $c_1$} & ${\bf {\scriptstyle 0.47}}$  &  ${\bf {\scriptstyle 0.48}}$  &  ${\bf {\scriptstyle 0.45}}$  &  ${\bf {\scriptstyle 0.43}}$ \\
{\large $c_2$} & ${\bf {\scriptstyle 0.83}}$ & ${\bf {\scriptstyle 0.94}}$ & ${\bf {\scriptstyle 0.80}}$ & ${\bf {\scriptstyle 0.90}}$ \\
{\large $c_3$} & ${\bf {\scriptstyle 0.11}}$  & ${\bf {\scriptstyle 0.15}}$ & ${\bf {\scriptstyle 0.14}}$ & ${\bf {\scriptstyle 0.21}}$ \\
{\large $c_4$} & ${\bf {\scriptstyle 0.84}}$ & ${\bf {\scriptstyle 0.83}}$ & ${\bf {\scriptstyle 0.87}}$ & ${\bf {\scriptstyle 0.87}}$ \\
{\large $c_5$} & ${\bf {\scriptstyle -0.21}}$ & ${\bf {\scriptstyle -0.20}}$ & ${\bf {\scriptstyle -0.28}}$ & ${\bf {\scriptstyle -0.28}}$ \\
{\large $c_6$} & --- & --- & ${\bf {\scriptstyle 0.023}}$ & ${\bf {\scriptstyle 0.023}}$ \\
{\large $c_7$}$\equiv \ln M_0$ & ${\bf {\scriptstyle 27.3}}$ & ${\bf {\scriptstyle 27.4}}$ & ${\bf {\scriptstyle 27.4}}$ & ${\bf {\scriptstyle 27.5}}$
\end{tabular}
\end{table*}

To gain an  idea of the acceptable analytical form for $\Gamma(z)$ and $f_{\rm duty}(z)$, we initially allow these quantities to have free values in all of the redshift bins: $\Gamma(z_i)$, $f_{\rm duty}(z_i)$, $i=1\ldots 9$. Thus, including $M_0$ we have $19$ free parameters in total. Figure~\ref{fig1} shows best fit values along with $1\sigma$ regions for $\Gamma(z_i)$ and $f_{\rm duty}(z_i)$ obtained from out MCMC calculations. We see that $\Gamma(z_i)$ are quite well fitted by a simple linear form.

As a next step, we fix $\Gamma(z)$ to have a linear redshift dependence $\Gamma(z)=c_4+c_5z$ and allow $f_{\rm duty}(z_i)$ to remain free parameters, so our free parameters are $M_0$, $c_4$, $c_5$, $f_{\rm duty}(z_i)\,(i=1\ldots 9)$, i.e. 12 in total. The best fit values for $f_{\rm duty}(z_i)$ are shown with red crosses in the upper panel of Fig.~\ref{fig1}. It turns out that these points can be approximated reasonably well with the analytic form,
\begin{equation}\label{eq3}
f_{\rm duty}(z)=\frac{c_1^2z}{(z-c_2)^2+c_1^2}+c_3\,,
\end{equation}
i.e., a Lorentzian profile multiplied by $z$ and a constant offset $c_3$ added.\footnote{In our earlier paper \citet{2012A&A...547A..21H}, where we also used the MF-to-LF mapping in the form of Eq.~(\ref{eq2}), we had a different analytic form for $f_{\rm duty}(z)$, which turns out to give somewhat poorer fit to the observed LF.}

To check for the self-consistency of the above analytic description we now fix $f_{\rm duty}(z)$ to have form as given in Eq.~(\ref{eq3}), but allow the power law indices in all nine redshift bins to be free parameters, i.e., in this case we have 13 free parameters: $M_0$, $c_1$, $c_2$, $c_3$, $\Gamma(z_i)\,(i=1\ldots 9)$. The best fit values for $\Gamma(z_i)$ are plotted as red crosses in the lower panel of Fig.~\ref{fig1}, where with solid lines we also show the best linear and quadratic approximations. Even though the linear approximation for $\Gamma(z)$ seems to capture the redshift dependence of the power law index quite well, the quadratic form with its one extra parameter, as it turns out, is statistically well justified.

Thus, our analytic MF-to-LF mapping is given by Eqs.~(\ref{eq2}), (\ref{eq1}), and (\ref{eq3}). In the following we call the models with linear and quadratic approximation for $\Gamma(z)$  Model I  and Model II, respectively. We considered models with $f_{\rm duty}$ constrained by the condition $f_{\rm duty}\le 1$, as well as unconstrained models, with $f_{\rm duty}$ allowed to take any positive values. Obviously, constrained models correspond to the case when there is at most one AGN per dark matter halo which is turned on with a probability given by $f_{\rm duty}$, whereas in the unconstrained models an arbitrary number of AGN per dark matter halo is allowed. It is interesting that in unconstrained models, $f_{\rm duty}$ does not reach values  much above unity, as illustrated by  the shaded region in the upper panel of Fig.~\ref{fig1}.  

\begin{table}
\centering
\caption{Overall performance of various fitting forms.}
\label{tab2}
\begin{tabular}{l|c|c|c|c|}
{\scriptsize Model} & {\scriptsize\# param} & {\scriptsize \# dof} & {\scriptsize $\chi^2$} & {\scriptsize expected $\chi^2$} \\
      &              &       &          & {\scriptsize (1$\sigma$ \emph{CL})} \\
\hline
{\scriptsize PLE} & $7$ & $72$ & $96.2$ & $72 \pm 12.0$\\
{\scriptsize LADE} & $8$ & $71$ & $67.4$ & $71 \pm 11.9$\\
{\scriptsize LDDE} & $9$ & $70$ & $57.8$ & $70 \pm 11.8$\\
\hline
{\scriptsize Model I} & $6$ & $73$ & $77.2$ & $73 \pm 12.1$\\
{\scriptsize Model I} {\tiny $+f_{\rm duty}\le 1$}& $6$ & $73$ & $83.4$ & $73 \pm 12.1$\\
\hline
{\scriptsize Model II} & $7$ & $72$ & $71.6$ & $72 \pm 12.0$\\
{\scriptsize Model II} {\tiny $+f_{\rm duty}\le 1$} & $7$ & $72$ & $77.0$ & $72 \pm 12.0$
\end{tabular}
\end{table}

In Table~\ref{tab1} we show the best fitting values for all the parameters of Models I and II, with and without an additional constraint on $f_{\rm duty}$. The calculated $\chi^2$ values along with expectations are given in Table~\ref{tab2}, where we also show results for empirical models used to describe observed X-ray LFs, namely  pure luminosity evolution (PLE), luminosity and density evolution (LADE), and luminosity-dependent density evolution (LDDE) models, with their analytic forms taken from \citet{2010MNRAS.401.2531A}. To perform a fair comparison between different analytic fitting forms, we have recalculated the best fit models for the latter using the same fitting machinery as for our models. It can be seen that, except for PLE, all of those models provide statistically valid descriptions of the observational data. Even though our MF-to-LF mapping does not provide as good a fit as LADE or LDDE models, it is quite remarkable that this more physically motivated model indeed seems to work. As a bonus, once MF-to-LF mapping is fixed, our models make several other predictions, which are discussed in the next section.

To illustrate resulting parameter uncertainties, we show in Fig.~\ref{fig2} $1\sigma$ and $2\sigma$ marginalized error regions along with 1D probability distribution functions for all the seven parameters of unconstrained Model II. The black dots and red crosses mark the maximum likelihood points for unconstrained and constrained (i.e., $f_{\rm duty}\le 1$) cases, respectively. We see that all the model parameters seem to be sufficiently well determined. The strongest degeneracy is seen between parameters $c_5$ and $c_6$, which might leave some doubts about the usefulness of the quadratic term in $\Gamma(z)$. However, as seen from the results shown in Table~\ref{tab2}, the need for parameter $c_6$ is statistically justified, at a confidence level corresponding to  $\sim 2.5\sigma$.

In Fig.~\ref{fig3} we plot the LFs in nine redshift bins for the best fitting PLE, LADE, LDDE models, and for our constrained Model II. Original data points with error bars, as determined by \citet{2010MNRAS.401.2531A}, are also shown. Our model slightly but systematically overshoots the highest luminosity data point at high redshifts. This is discussed in more detail in the following section.

\begin{figure*}
\centering
\includegraphics[width=\plotwdtwo]{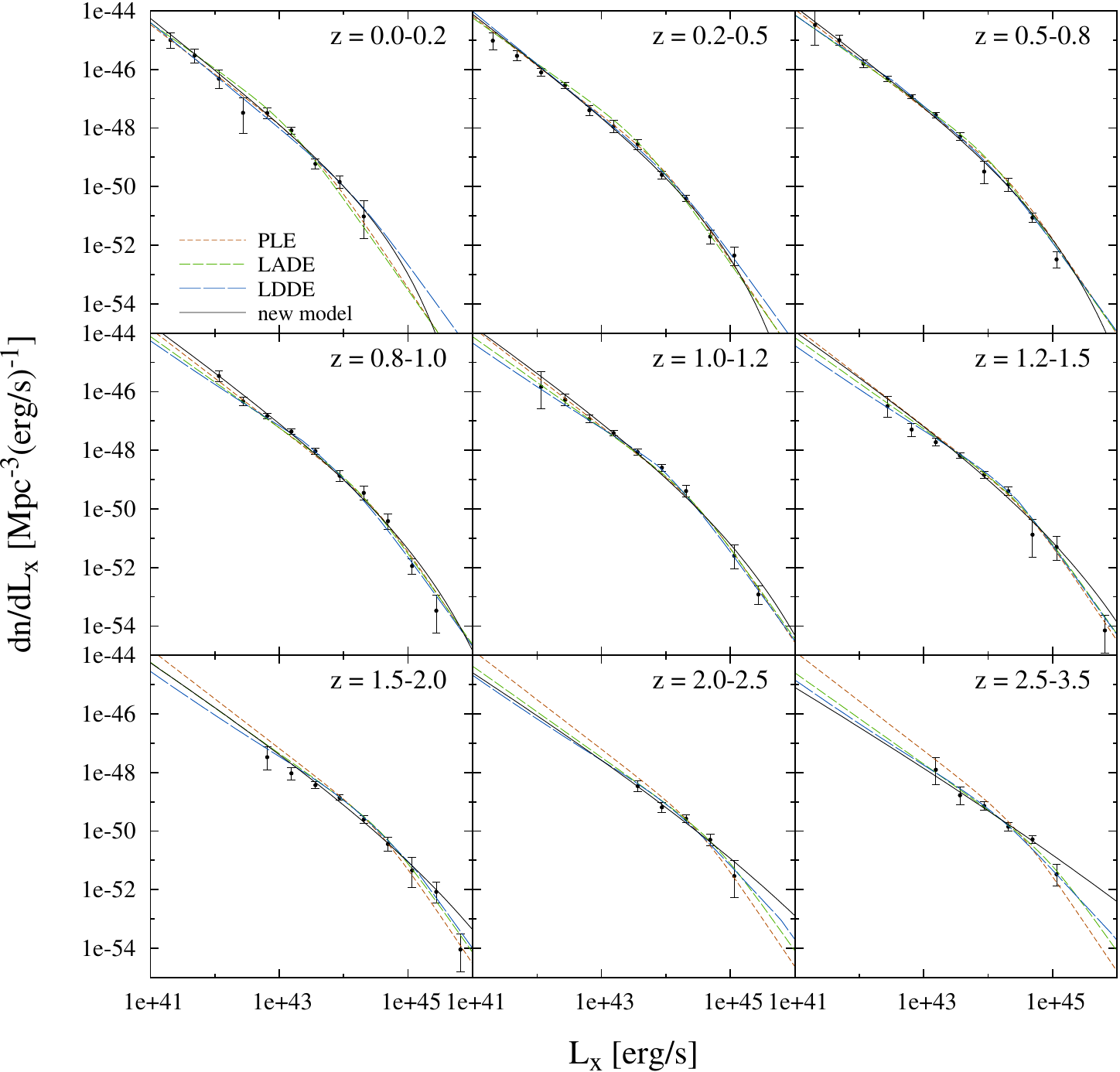}
\caption{Best fitting X-ray AGN LFs in nine redshift bins along with observational data from \citet{2010MNRAS.401.2531A}. In addition to our constrained Model II fits we also show the best fitting PLE, LADE, and LDDE models with the corresponding analytic forms taken from \citet{2010MNRAS.401.2531A}.}
\label{fig3}
\end{figure*}

\section{Implications and shortcomings of the model}\label{sec3}

In this section we investigate some of the consequences of the MF-to-LF mapping models presented above. 

\subsection{Comoving AGN density}

\begin{figure}
\centering
\includegraphics[width=\plotwd]{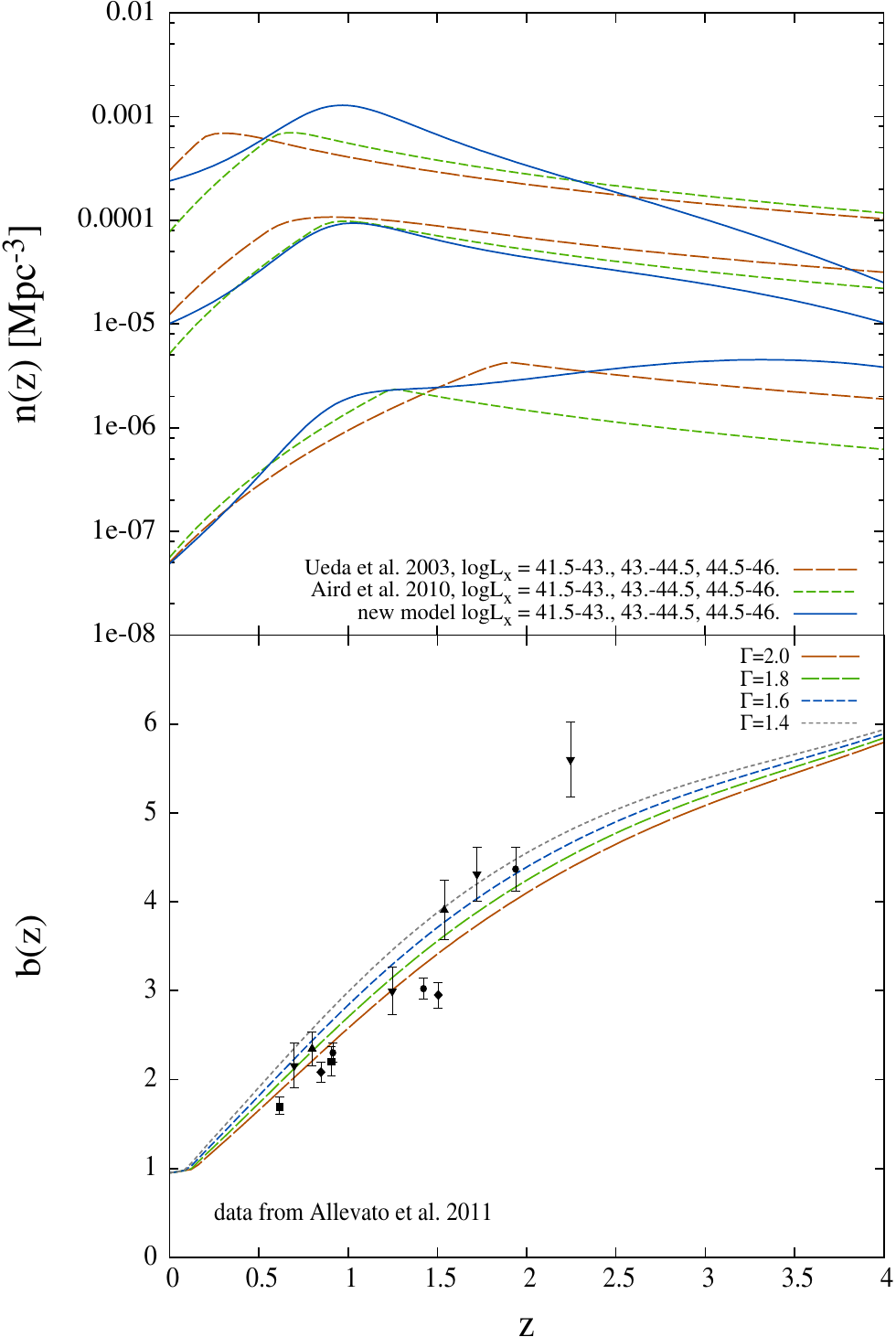}
\caption{\emph{Upper panel:} comoving X-ray AGN number densities as a function of redshift for three $2-10$ keV band luminosity ranges as shown in the legend. The luminosity is increasing from top to bottom. Along with our constrained Model II we have also plotted the results arising from the LF models of \citet{2003ApJ...598..886U} and \citet{2010MNRAS.401.2531A}. \emph{Lower panel:} comparison of bias parameters from the constrained Model II with the observational data from \citet{2011ApJ...736...99A}. The four lines assume different effective photon indices ($\Gamma=1.4,1.6,1.8,2.0$) for the approximate transformation between soft and hard X-ray bands.}
\label{fig4}
\end{figure}

We start by investigating the comoving number density of hard-band selected X-ray AGN and its dependence on redshift. In the upper panel of Fig.~\ref{fig4} we have plotted comoving number density for three X-ray luminosity intervals as specified in the legend. Here the highest (lowest) group of curves correspond to the lowest (highest) luminosities. Along with constrained Model II results, for comparison we have also plotted number densities resulting from \citet{2003ApJ...598..886U} and \citet{2010MNRAS.401.2531A} LF fits. In the case of \citet{2003ApJ...598..886U} we directly use their best fit LF parameters, whereas for the  \citet{2010MNRAS.401.2531A} LDDE model, we used best fit parameters obtained in the previous section.
The comparison of our results with \citet{2010MNRAS.401.2531A} LDDE parameterization should give some idea about the possible level of uncertainty in the so-obtained $n(z)$, since both of the models give statistically valid fits to LF data. 

This type of figure is often used to illustrate the antihierarchical growth of SMBHs (as manifested by the AGN activity) where the peak in number density moves toward lower redshifts as one considers intrinsically less luminous AGN. 
However,  at least in the case of  \citet{2010MNRAS.401.2531A} LF data, the conclusion about the behavior of the maximum in $n(z)$ curve may depend on the particular choice of the functional form used to fit the X-ray LF data.
Indeed, in our LF model, the $n(z)$ maxima for the lowest luminosity bins stay almost at the same position, while for the highest luminosity bin we see gradual movement toward higher redshifts. This does not seem to be entirely incompatible with the pattern of the original LF data points in Fig. 10 in  \citet{2010MNRAS.401.2531A}. Furthermore, in our model the rate of change of $n(z)$ has quite a strong dependence on luminosity, with the lowest luminosity AGN having the fastest rise in number density going from high redshifts down to $z\sim 1$,  the AGN in the intermediate luminosity bin having only very mild increase and the highest luminosity bin showing a mild drop in the volume density with the redshift. The latter is related to the model behavior at high luminosity and redshift, discussed below.

\subsection{A possible shortcoming of the model}
\label{sec:shorts}

A careful examination of Fig.~\ref{fig3} reveals that our model overpredicts the numbers of most luminous objects at high redshift (see the three lower panels in Fig.~\ref{fig3}). This may point at the shortcoming of the model. Such a shortcoming could be caused by several reasons. One possible cause is the lack of proper treatment of the Eddington luminosity limit. The Eddington ratio analysis later on in this section suggests that this factor may become important in the halo mass  range $\log M_{\rm h}\ga 13.5-14.5$ at the redshift $z\ga 2-3$.  On the other hand, it may also be an indication  of the shortcoming of the underlying halo mass function, somewhat overpredicting the abundance of massive halos at high redshifts. 

Another possible reason may be  a failure of the simple power-law scaling between the luminosity and halo mass, Eq.~(\ref{eq1}), at high halo masses and redshifts. Indeed, due to obvious considerations of time available for the  black hole growth,  the most massive halos at high redshifts may harbor somewhat less massive black holes than implied by Eq.~(\ref{eq1}). The parameters of the halo mass -- luminosity scaling  were determined from the global LF fit that is determined by the bulk of the LF data points and is not very sensitive to the data at the extremities because
the contribution of high luminosity points to $\chi^2$ is relatively small, owing to their small number and  also to their relatively larger uncertainties.  

Despite this shortcoming, however, the model provides a statistically valid overall description of the X-ray LF data (Table \ref{tab2}).  Its (possible) failure  at the high luminosity end only affects very few of the most massive halos at high redshifts and does not diminish the overall predictive power of the model for  the bulk of typical AGN, as discussed in the following sections.

\subsection{AGN clustering}

Since our model has a fixed scaling relation between the halo mass and AGN X-ray luminosity, it can straightforwardly predict the AGN clustering strength.Indeed, the AGN clustering bias can be computed as
\begin{equation}\label{eq1new}
b(z)=\frac{\int b_{\rm h}[M_{\rm h}(L_{\rm X},z),z]\frac{{\rm d}n}{{\rm d}L_{\rm X}}(L_{\rm X},z)\,{\rm d}L_{\rm X}}{\int \frac{{\rm d}n}{{\rm d}L_{\rm X}}(L_{\rm X},z)\,{\rm d}L_{\rm X}}\,,
\end{equation}
where the dark matter clustering bias is taken from the analytical model of \citet{2001MNRAS.323....1S}, and  the $L_{\rm X}$ integration bounds are appropriately adjusted to take the details of how the X-ray AGN sample was selected into account. In particular, for a flux-limited sample $L_{{\rm X},\min}(z)=4\pi D_L^2(z)F_{\rm X,lim}$, where $D_L(z)$ is the luminosity distance to redshift $z$.

In the lower panel of Fig.~\ref{fig4} we compare the thus-computed clustering bias as a function of redshift with the AGN clustering data from \citet{2011ApJ...736...99A}. \citet{2011ApJ...736...99A} used soft band X-ray data from COSMOS field, which covers $2.13$ deg$^2$ and contains a total of 780 AGN with available redshifts, its spectroscopic completeness being $\sim 53\%$. In their analysis \citet{2011ApJ...736...99A} applied a magnitude cut $I_{\rm AB}<23$ and redshift cut $z<4$, resulting in 593 objects in their final sample, with effective spectroscopic completeness  of $\sim 65\%$.

Since the clustering bias in our model depends on AGN luminosity, with more luminous AGN having higher bias parameters, we have to adjust the effective flux limit appropriately in order to make a fair comparison. Since the selection of \citet{2011ApJ...736...99A} sample was done in the soft band, we  transformed their flux limit to the corresponding limit in the hard band. We do this in a simple way by assuming an effective population-averaged spectral index $\Gamma$.\footnote{Not to be confused with the power law index $\Gamma(z)$ appearing in Eq.~(\ref{eq1}).} The results for four different values of $\Gamma$ are shown in the lower panel of Fig.~\ref{fig4}. Because in the hard band one starts to see more and more sources that are obscured in the soft band, the effective spectral index around $\sim 1.6$ appears to be the most appropriate one. Considering this, we see that except for the highest redshift bin, our predictions for the bias parameter  agree reasonably well with the measurements. The highest redshift bin is represented by one point only, and it remains to be seen how significant the deviation is. One has to keep in mind that the measurement of the linear bias parameter, and especially the proper estimation of the errors due to cosmic variance (e.g., accidental superclusters in small fields could significantly influence the results) from the fields as small as $\sim 2$ deg$^2$, may be  quite a cumbersome task.

\subsection{AGN duty cycle}

In both constrained and unconstrained models, the AGN duty cycle (Eq.~(\ref{eq2})) peaks at the redshift $z\simeq 1$ (Fig.~\ref{fig1}). Although unconstrained models (when the duty cycle was allowed to take any positive value) result in a somewhat better description of the observed X-ray LF data, it is interesting that the peak value of the duty cycle in these models is nevertheless close to unity, with the upper limit of $f_{\rm duty} \la 1.5$ ($1\sigma$ confidence). This implies that on average there is at most $\sim$one AGN per dark matter halo.  

The  value for the duty cycle close to unity is a consequence of the fact that at the redshift $z\sim 1$, the volume density of X-ray AGN is comparable to the volume density of the dark matter halos. Indeed, a power-law $M_{\rm h}$-$L_{\rm X}$ scaling relation  does not introduce any new scale, therefore the only way to obtain agreement with the observed LF is by matching the MF characteristic mass $M_*$ with a characteristic AGN X-ray luminosity $L_*$.  In this MF to LF mapping, the role of the $f_{\rm duty}$ is to tally their volume densities. As it turns out, the comoving number densities of $M_*$ (see Appendix~\ref{appa}) and $L_*$ objects are comparable at $z\sim 1$: $\left[L_{\rm X}\frac{{\rm d}n}{{\rm d}L_{\rm X}}(L_{\rm X}=L_*,z\simeq 1)\simeq M_{\rm h}\frac{{\rm d}n}{{\rm d}M_{\rm h}}(M_{\rm h}=M_*,z\simeq 1)\right]$; i.e., there is approximately as many dark matter halos as AGN. Correspondingly, $f_{\rm duty}\sim 1$. The duty cycle drops to $f_{\rm duty}\sim 0.2$ at the redshift $z=0$ (Fig.~\ref{fig1}), a manifestation of the well known paucity of AGN in the local Universe. It is interesting to note that the duty cycle at $z=0$ is not as low as predicted by the simple continuity equation models with a fixed Eddington ratio, but is closer to the ones in which the median of the Eddington ratio distribution decreases with redshift.

\subsection{$M_{\rm h}-L_{\rm X}$ scaling relation} 

\begin{figure}
\centering
\includegraphics[width=\plotwd]{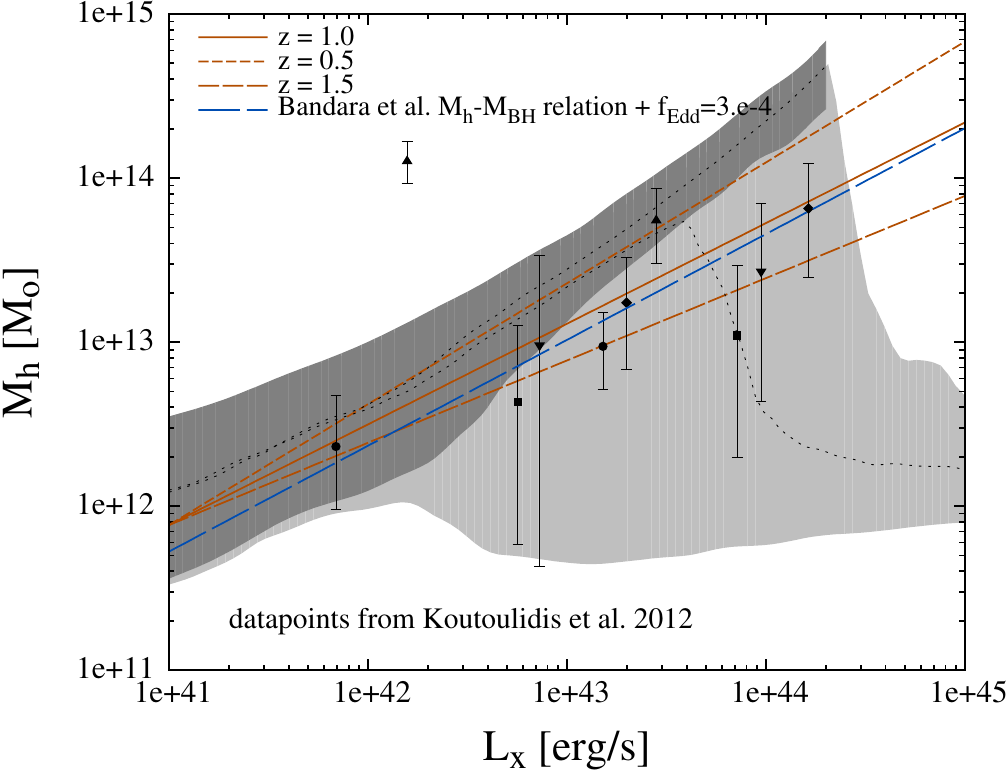}
\caption{Comparison of the $M_{\rm h}$-$L_{\rm X}$ scaling relation with the observational data from \citet{2012MNRAS.tmp..101K}. To capture the redshift range of majority of AGN used in this analysis, we have plotted the model scaling relations for three redshifts: $z=0.5,1.0,1.5$. The dotted lines with surrounding gray bands display the results from the semianalytic galaxy-AGN co-evolution model of \citet{2012MNRAS.419.2797F}, as presented in \citet{2012MNRAS.tmp..101K}, showing their `hot-halo' (dark gray region) and `all AGN' case, i.e., the sum of `hot-halo' and `starburst' modes (light gray region). The long-dashed line shows the $M_{\rm h}$-$L_{\rm X}$ scaling one obtains once converting the amplitude-adjusted $M_{\rm h}$-$M_{\rm BH}$ relation of \citet{2009ApJ...704.1135B} by using the fixed Eddington ratio $f_{\rm Edd}\equiv L_{\rm X}/L_{\rm Edd}(M_{\rm BH})=3\times10^{-4}$.}
\label{fig5}
\end{figure}

As emphasized above, the crucial component in our model is the assumption about the existence of  a scaling relation between the halo mass and AGN X-ray luminosity.  \citet{2012MNRAS.tmp..101K} have recently attempted to infer this type of scaling by using AGN clustering data from several deep X-ray fields: Chandra Deep Field South and North, the AEGIS, the extended Chandra Deep Field South, and the COSMOS field. In Fig.~\ref{fig5} we show their measurements with error bars along with our constrained Model II predictions for redshifts $0.5$, $1.0$, and $1.5$, i.e. approximately covering the redshift range of AGN used in their analysis. The dotted lines with surrounding gray bands display the results from the semianalytic galaxy-AGN co-evolution model of \citet{2012MNRAS.419.2797F} (as presented in Fig. 9 of \citet{2012MNRAS.tmp..101K}), showing their `hot-halo' (dark grey region) and `all AGN' case, i.e., the sum of `hot-halo' and `starburst' modes (light gray region). Although the uncertainties are rather large, we see that our model is able to capture the major trend where the more luminous AGN tend to populate more massive halos. The quantitative agreement between our $M_{\rm h}-L_{\rm X}$ relation and  results of \citet{2012MNRAS.tmp..101K} is truly  remarkable since they are based on totally independent arguments.

\begin{figure}
\centering
\includegraphics[width=\plotwd]{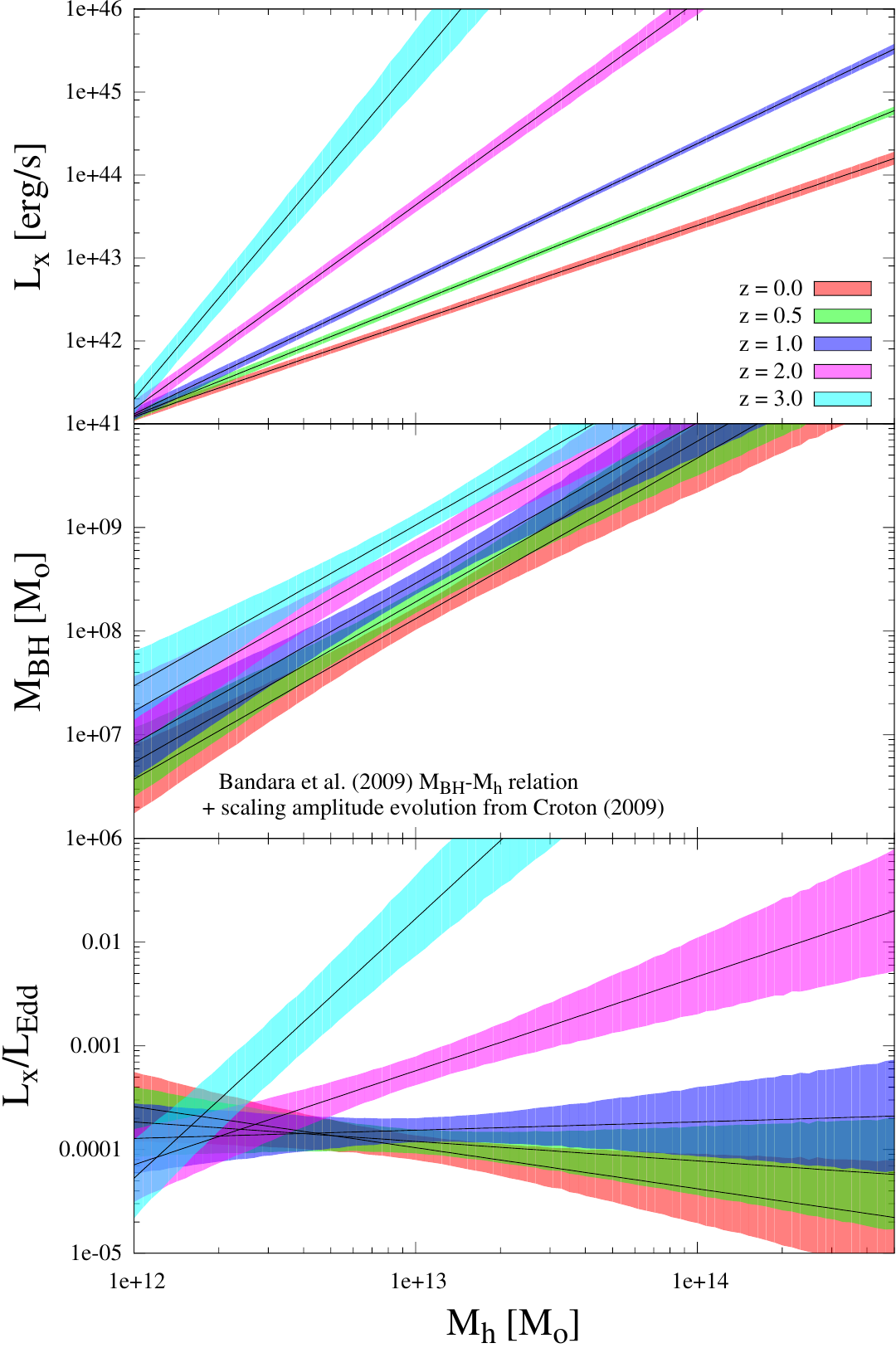}
\caption{\emph{Upper panel:} average $L_{\rm X}$-$M_{\rm h}$ relations (Eq.~(\ref{eq1})) along with $1\sigma$ uncertainties for five different redshifts as given in the legend. \emph{Middle panel:} amplitude-adjusted $M_{\rm BH}$-$M_{\rm h}$ scaling relation with $1\sigma$ uncertainty from \citet{2009ApJ...704.1135B}. \emph{Lower panel:} Eddington ratios in $2-10$ keV band for five redshifts arising from $L_{\rm X}$-$M_{\rm h}$ and $M_{\rm BH}$-$M_{\rm h}$ relations given in the upper two panels.}
\label{fig6}
\end{figure}

The long-dashed line in Fig.~\ref{fig5} shows the $M_{\rm h}$-$L_{\rm X}$ scaling one obtains when converting the $M_{\rm h}$-$M_{\rm BH}$ relation of \citet{2009ApJ...704.1135B} by assuming a constant Eddington ratio $f_{\rm Edd}\equiv L_{\rm X}/L_{\rm Edd}(M_{\rm BH})$ of $3\times 10^{-4}$. In reality, because the \citet{2009ApJ...704.1135B} $M_{\rm BH}$-$M_{\rm h}$ scaling relation used strong lensing masses derived from the SLACS lens sample with mean redshift $z\simeq 0.2$, we have adjusted the amplitude of the scaling relation to the redshift $z\sim 1$, using results from \citet{2009MNRAS.394.1109C} semi-analytic models. In \citet{2009ApJ...704.1135B} the authors obtain $M_{\rm BH} \propto M_{\rm h}^{\gamma}$ with $\gamma=1.55\pm 0.31$, which agrees quite well with the analytic and semi-analytic models of \citet{2003ApJ...595..614W} and \citet{2009MNRAS.394.1109C}, which in turn provide values $\gamma=5/3\simeq 1.67$ and $1.39$, respectively. In \citet{2009MNRAS.394.1109C} and \citet{2003ApJ...595..614W} models the amplitude of the scaling relation increases by factors of $\sim2$ and $\sim4$, respectively, while going from $z=0$ to $z=1$. However, the power law index $\gamma$ is independent of redshift in both models.

\subsection{AGN Black hole masses, Eddington ratios, and the growth of SMBHs}

By assuming the $M_{\rm BH}$-$M_{\rm h}$ scaling of \citet{2009ApJ...704.1135B} with amplitude adjusted according to \citet{2009MNRAS.394.1109C} semi-analytic model, we can convert our $L_{\rm X}$-$M_{\rm h}$ relation to $f_{\rm Edd}$-$M_{\rm h}$ scaling relation. This is detailed in Fig.~\ref{fig6} where we have plotted various scalings for five different redshifts. The upper panel gives our results for $L_{\rm X}$-$M_{\rm h}$ with $1\sigma$ uncertainties as determined in Section~\ref{sec2} for the constrained Model II. The middle panel shows the amplitude-adjusted \citet{2009ApJ...704.1135B} $M_{\rm BH}$-$M_{\rm h}$ relation with $1\sigma$ uncertainties. In the lower panel we show the Eddington ratios  as determined by combining the results from the two uppermost panels. The X-ray luminosities and Eddington ratios refer to the 2--10 keV band and no bolometric corrections were applied. For a typical type I AGN spectrum, the bolometric correction factor should be in the $\sim 5-20$ range \citep{2012MNRAS.425..623L}.

As we can see, the Eddington ratios for high redshift ($z\gtrsim 2$) AGN are predicted to be steeply rising functions of halo mass, while at redshifts $z\sim 1$, i.e. close to the peak in X-ray AGN number density, the dependence of $f_{\rm Edd}$ on $M_{\rm h}$ flattens out, and at lower redshifts the trend is mildly reversed; i.e., the AGN in less massive halos have somewhat higher Eddington ratios. For $z\sim 3$,  the 2--10 keV band $f_{\rm Edd}$ reaches values of $\sim 0.1$ for halo masses $\sim 2\times 10^{13}\,M_\odot$; i.e. the bolometric Eddington ratio will approach $\sim1$ at this mass. At higher masses, our $M_{\rm h}-L_{\rm X}$ scaling relation  would imply $f_{\rm Edd}\ga 1$. Obviously, this is because our simple $M_{\rm h}-L_{\rm X}$ scaling relation does not impose the Eddington luminosity limit. This  may be considered as one of the shortcomings of the model, as discussed in section \ref{sec:shorts}. One has to note, however, that this does not affect the overall performance of the model, in the major part of the parameter space. Indeed, such massive halos at redshifts $z\sim 3$ are extremely rare, corresponding to $\sim 4-5\sigma$ peaks in the initial fluctuation field. The bulk of the objects are shining at much lower Eddington ratios, $\ll 1$.

\begin{figure}
\centering
\includegraphics[width=\plotwd]{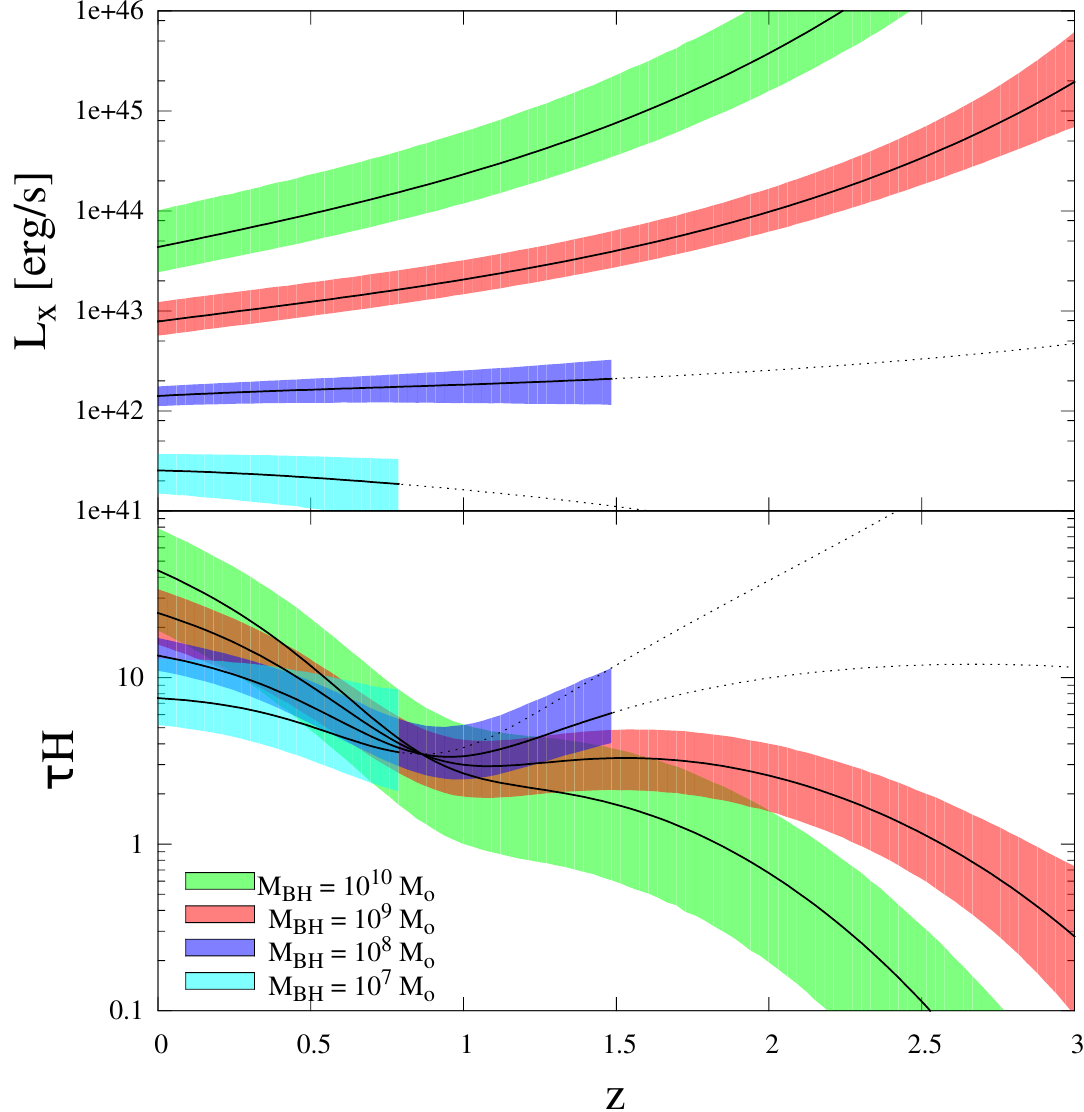}
\caption{The hard band X-ray luminosities (\emph{upper panel}) and SMBH e-folding times $\tau\equiv \frac{M_{\rm BH}}{\dot{M}_{\rm BH}}$ in units of the Hubble time (\emph{lower panel}) for various black hole masses as shown in the legend. For the low mass SMBHs at higher redshifts, where they are too faint to be constrained by the observational data, we have shown the extrapolations of the model curves with dotted lines and have omitted the corresponding $1\sigma$ uncertainty bands.}
\label{fig7}
\end{figure}

Having specified $M_{\rm BH}$-$M_{\rm h}$ scaling relation, we can also calculate how SMBHs with different masses grow throughout the cosmic time. Taking the duty cycle into account, the population-averaged e-folding time of the SMBH growth is
\begin{equation}
\tau\equiv \frac{M_{\rm BH}}{\dot{M}\, f_{\rm duty}}=\frac{\eta\, M_{\rm BH}c^2 }{f_{\rm duty}\,f_{\rm bol}\, L_{\rm X}}
\end{equation}
where $\eta$ is the accretion efficiency, defined as 
\begin{equation}
L_{\rm X}=f_{\rm bol}^{-1}\, \eta\,\dot{M}c^2
\end{equation}
$L_{\rm X}$ refers to the X-ray luminosity in the 2--10 keV band used throughout this paper, and $f_{\rm bol}\sim 5-20$ is the bolometric correction factor for the 2--10 keV band. The e-folding time can be expressed further though the Eddington ratio:
\begin{equation}
\tau=\frac{\eta\, M_{\rm BH}c^2 }{f_{\rm duty}\,f_{\rm bol}\,f_{\rm Edd}\, L_{\rm Edd}}=
0.45\, \frac{\eta}{f_{\rm duty}\,f_{\rm bol}\,f_{\rm Edd}} {\rm ~ Gyrs}\,.
\end{equation}
In the lower panel of Fig.~\ref{fig7} we plot the e-folding time in units of the Hubble time, i.e. $\tau H(z)$, vs redshift for several values of the black hole mass. Shown in the upper panel is the evolution of the population average $2-10$ keV band X-ray luminosity with redshift. AGN harboring black holes with masses less than a $\sim$few$\times 10^8\,M_\odot$ are too faint at high redshifts and fall outside the luminosity range of the X-ray LF data of \citet{2010MNRAS.401.2531A} used in this study. Therefore they are not directly constrained by the X-ray data. At these redshifts we only show with dotted lines the extrapolations of the central model curves and omit the $1\sigma$ uncertainty regions.

We see that the X-ray luminosity of  more massive SMBHs declines faster with the redshift than for less massive ones.  For $\sim 10^8\,M_\odot$ black holes, $L_{\rm X}$ stays almost constant, while for even lower mass SMBHs, the evolutionary trend is slightly reversed; i.e., the objects are getting somewhat brighter as one moves toward lower redshifts. As one might have expected from the results plotted in the bottom panel of Fig.~\ref{fig6}, for the typical X-ray-selected AGN, the corresponding SMBH e-folding times are long in comparison to the Hubble time. With the exception of most massive SMBHs at high redshifts, the typical X-ray-selected AGN are not in the rapid mass growth regime in the $z\sim 0-3$ redshift range.

The crossing of the model curves in the lower panel of Fig.~\ref{fig7} at the  redshift of $z\sim 0.85$ is artificial and does not seem to have any physical meaning. It is driven by the equality of the power law indices of the $M_{\rm BH}$-$M_{\rm h}$ and $M_{\rm h}-L_{\rm X}$ scaling relations, which leads to the $\tau H$ being independent of $M_{\rm BH}$ at that particular $z$.

\begin{table*}
\centering
\caption{Best fitting parameter values and $\chi^2$ values for various models, discussed in detail in the main text.}
\label{tab3}
\begin{tabular}{l|l|l|l|l|l|l|l|l|l|l|l|}
& {\large $c_1$} & {\large$c_2$} & {\large $c_3$} & {\large $c_4$} & {\large $c_5$} & {\large $c_6$} &  {\large $c_7$}$\equiv \ln M_0$ &{\large $c_8$}$\equiv \sigma$ & {\large $c_9$}$\equiv \alpha$ & \# dof & $\chi^2$\\
\hline
$f_{\rm duty}(z)$ \& free scatter & ${\bf {\scriptstyle 0.425}}$ & ${\bf {\scriptstyle 0.892}}$ & ${\bf {\scriptstyle 0.219}}$ & ${\bf {\scriptstyle 0.870}}$ & ${\bf {\scriptstyle -0.283}}$ & ${\bf {\scriptstyle 0.0252}}$ & ${\bf {\scriptstyle 27.5}}$ & ${\bf {\scriptstyle 0.0458}}$ & ${\bf {\scriptstyle 0.}}$ (fix) & ${\bf {\scriptstyle 71}}$ & ${\bf {\scriptstyle 71.6}}$\\
$f_{\rm duty}(z)$ \& 0.25 dex scatter & ${\bf {\scriptstyle 0.445}}$ & ${\bf {\scriptstyle 0.907}}$ & ${\bf {\scriptstyle 0.195}}$ & ${\bf {\scriptstyle 0.893}}$ & ${\bf {\scriptstyle -0.291}}$ & ${\bf {\scriptstyle 0.0227}}$ & ${\bf {\scriptstyle 27.6}}$ & ${\bf {\scriptstyle 0.25\ln 10}}$ (fix) & ${\bf {\scriptstyle 0.}}$ (fix) & ${\bf {\scriptstyle 72}}$ & ${\bf {\scriptstyle 75.7}}$\\
$f_{\rm duty}(z)$ \& 0.50 dex scatter & ${\bf {\scriptstyle 0.485}}$ & ${\bf {\scriptstyle 0.936}}$ & ${\bf {\scriptstyle 0.157}}$ & ${\bf {\scriptstyle 0.936}}$ & ${\bf {\scriptstyle -0.295}}$ & ${\bf {\scriptstyle 0.0148}}$ & ${\bf {\scriptstyle 27.8}}$ & ${\bf {\scriptstyle 0.50\ln 10}}$ (fix) & ${\bf {\scriptstyle 0.}}$ (fix) & ${\bf {\scriptstyle 72}}$ & ${\bf {\scriptstyle 81.6}}$\\
$f_{\rm duty}(L_{\rm X},z)$, no scatter & ${\bf {\scriptstyle 0.372}}$ & ${\bf {\scriptstyle 0.820}}$ & ${\bf {\scriptstyle 0.224}}$ & ${\bf {\scriptstyle 0.994}}$ & ${\bf {\scriptstyle -0.299}}$ & ${\bf {\scriptstyle 0.0319}}$ & ${\bf {\scriptstyle 27.0}}$ & ${\bf {\scriptstyle 0.}}$ (fix)& ${\bf {\scriptstyle 0.306}}$ & ${\bf {\scriptstyle 71}}$ & ${\bf {\scriptstyle 46.9}}$\\
$f_{\rm duty}(L_{\rm X},z)$ \& free scatter & ${\bf {\scriptstyle 0.389}}$ & ${\bf {\scriptstyle 0.831}}$ & ${\bf {\scriptstyle 0.201}}$ & ${\bf {\scriptstyle 1.20}}$ & ${\bf {\scriptstyle -0.389}}$ & ${\bf {\scriptstyle 0.0422}}$ & ${\bf {\scriptstyle 27.3}}$ & ${\bf {\scriptstyle 1.39}}$ & ${\bf {\scriptstyle 0.414}}$ & ${\bf {\scriptstyle 70}}$ & ${\bf {\scriptstyle 43.6}}$
\end{tabular}
\end{table*}

Thus, with the particular form of the $M_{\rm BH}$-$M_{\rm h}$ scaling relation from  \citet{2009ApJ...704.1135B}  with amplitude evolution from \citet{2009MNRAS.394.1109C}, our model implies relatively low   Eddington ratios, $f_{\rm Edd}\sim 10^{-4}-10^{-3}$.   In the redshift range $z\sim 0-3$, for a typical X-ray-selected AGN, accretion proceeds in the `hot-halo mode' characterized by a relatively low mass accretion rate and long e-folding time for SMBH mass growth. Obviously, this conclusion critically  depends on the assumed $M_{\rm BH}$-$M_{\rm h}$ scaling relation. Indeed, as X-ray luminosity for the given halo mass is directly determined by our model from the $L_X-M_h$ mapping,  the Eddington ratio scales inversely with the black hole mass, while e-folding time is linearly proportional to the black hole mass.

\section{Extensions of the model}\label{newsec}

\begin{figure*}
\centering
\includegraphics[width=\plotwdtwo]{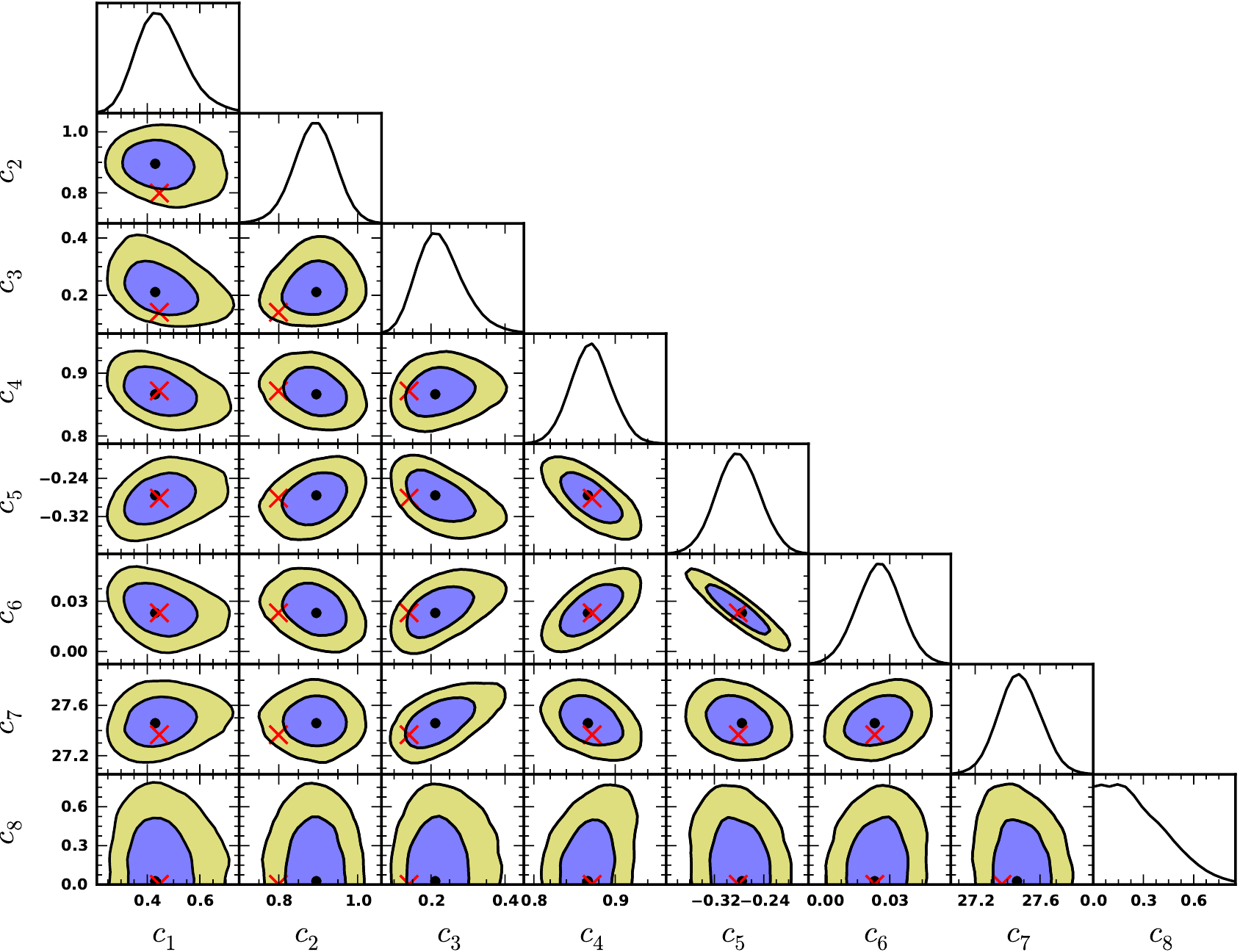}
\caption{Analog of Fig.~\ref{fig2} with one additional parameter, $c_8\equiv\sigma$, describing the scatter in the lognormal distribution.}
\label{fig8}
\end{figure*}

\begin{figure*}
\centering
\includegraphics[width=\plotwdtwo]{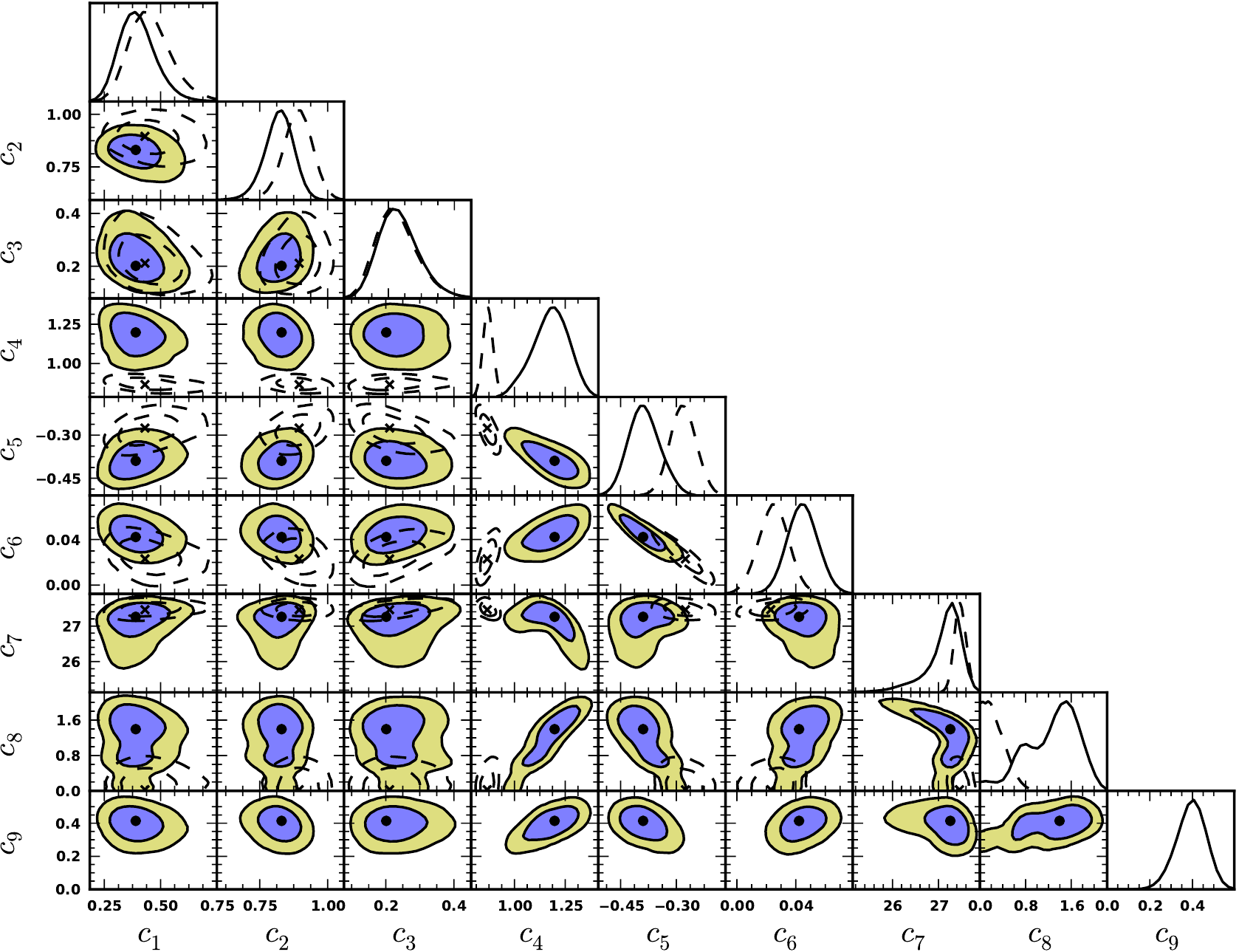}
\caption{Analog of Fig.~\ref{fig2} with two additional parameters, $c_8\equiv\sigma$ and $c_9\equiv\alpha$, representing the scatter parameter in the lognormal distribution and the power law index for the $L_{\rm X}$-dependent $f_{\rm duty}$, respectively. For comparison the contours from Fig.~\ref{fig8} are shown with dashed lines.}
\label{fig9}
\end{figure*}

The minimalistic  model considered in the previous sections has two major simplifications: a deterministic $M_{\rm h}-L_{\rm X}$ relation and a luminosity-independent duty cycle. In this section we lift these assumptions and explore a class of more sophisticated models.

We start with introducing scatter in the $M_{\rm h}-L_{\rm X}$ relation.  We assume that scatter can be described  by the lognormal probability distribution:\footnote{This assumption, along with the power law $M_{\rm BH}$-$M_{\rm h}$ scaling relation used throughout this paper, leads to an approximately lognormal probability distribution function for the Eddington ratio $f_{\rm Edd}$, which agrees quite well with the measurements of \citet{2012MNRAS.425..623L}. However, once the completeness correction is taken into account, at redshifts $z\lesssim 1$, the $f_{\rm Edd}$ distributions can become closer to a power-law shape, instead \citetext{see also~\citealp{2012ApJ...746...90A}}. Within our model framework the Eddington ratio distribution can be determined once fixing the $M_{\rm BH}$-$M_{\rm h}$ relation. In this paper we have assumed one particular form for the $M_{\rm BH}$-$M_{\rm h}$ scaling relation, i.e., a simple power law. In principle, there are quite large observational uncertainties regarding the particular form for this relation, and thus by allowing more freedom here, one should certainly be able to obtain a significantly better match with the claimed power law like $f_{\rm Edd}$ distributions at lower redshifts. However, this more detailed analysis is beyond the scope of the current work.}
\begin{multline}
f(L_{\rm X}|M_{\rm h},z)=\\
\frac{1}{L_{\rm X}\sigma\sqrt{2\pi}}\exp \left \{-\frac{\left [ \ln L_{\rm X} - \mu(M_{\rm h},z)\right ]^2}{2\sigma^2}\right \}\,,
\end{multline}
whose width  $\sigma$ is assumed to be independent of $M_{\rm h}$ and $z$.
Eq.~(\ref{eq2}) for the luminosity function  is now replaced by
\begin{multline}
\frac{{\rm d}n}{{\rm d}L_{\rm X}}(L_{\rm X},z)=\\
f_{\rm duty}(L_{\rm X},z)\int f(L_{\rm X}|M_{\rm h},z)\frac{{\rm d}n}{{\rm d}M_{\rm h}}(M_{\rm h},z)\,{\rm d}M_{\rm h}\,.
\label{eq:dndl_new}
\end{multline}

There are several obvious possibilities for $\mu(M_{\rm h},z)$, depending on whether one  chose the (i) mode, (ii) median, or (iii) mean of the lognormal probability distribution function to obey  our previous deterministic $M_{\rm h}-L_{\rm X}$ relation defined by eq.(\ref{eq1}). This gives 
\begin{eqnarray}
\mu(M_{\rm h},z)&=&\sigma^2 + \ln\left [ L_0\left ( \frac{M_{\rm h}}{M_0}\right )^{\frac{1}{\Gamma(z)}}\right ]\,,\label{eq4new}\\
\mu(M_{\rm h},z)&=&\ln\left [ L_0\left ( \frac{M_{\rm h}}{M_0}\right )^{\frac{1}{\Gamma(z)}}\right ]\,,\\
\mu(M_{\rm h},z)&=&-\frac{\sigma^2}{2} + \ln\left [ L_0\left ( \frac{M_{\rm h}}{M_0}\right )^{\frac{1}{\Gamma(z)}}\right ]\,
\end{eqnarray}
for the  mode, median, and mean, respectively. Since this analysis is largely  inspired by comparison with observations, which pick up the most probable objects, it is natural to choose the mode of the distribution, eq.(\ref{eq4new}).

It is easy to see that in case of small scatter, all the above descriptions are equivalent, and in the limit $\sigma \rightarrow 0$
\[
f(L_{\rm X}|M_{\rm h},z)\longrightarrow \delta\left [L-L_0\left ( \frac{M_{\rm h}}{M_0}\right )^{\frac{1}{\Gamma(z)}}\right ]\,,
\]
and we recover the deterministic relation  given in Eq.~(\ref{eq1}).

To calculate the AGN clustering bias, Eq.~(\ref{eq1new}) now generalizes to
\begin{equation}
b(z)=\frac{\int \int b_{\rm h}(M_{\rm h},z)f(L_{\rm X}|M_{\rm h},z)\frac{{\rm d}n}{{\rm d}M_{\rm h}}(M_{\rm h},z)\,{\rm d}M_{\rm h}{\rm d}L_{\rm X}}{\int \int f(L_{\rm X}|M_{\rm h},z)\frac{{\rm d}n}{{\rm d}M_{\rm h}}(M_{\rm h},z)\,{\rm d}M_{\rm h}{\rm d}L_{\rm X}}\,,
\end{equation}
Similar to eq.(\ref{eq1new}),  the integration limits for $L_{\rm X}$ have to be adjusted appropriately, taking the ways AGN sample was selected into account.

In Fig.~\ref{fig8}, which is the analog of the previous Fig.~\ref{fig2}, we show the results of introducing scatter in the $M_h-L_X$ mapping. The scatter is parametrized via a single parameter -- the width of the log-normal distribution  $\sigma\equiv c_8$. All other conventions are exactly the same as in Fig.~\ref{fig2}.  
The best fitting values for all the parameters are given in the first row of Table~\ref{tab3} for the case of unconstrained $f_{\rm duty}$. As is obvious from Fig.~\ref{fig8} and Table~\ref{tab3}, no scatter is required by the X-ray LF data, since the best-fit value of $\sigma$ is close to zero. Correspondingly, the $\chi^2$ and the best fit values of other parameters are nearly identical  to the ones obtained in the deterministic case (cf. Table~\ref{tab1}).
By comparing with the results shown in Fig.~\ref{fig2}, it is interesting to note that after allowing for the scatter in the $M_h-L_X$ relation,  no new strong parameter degeneracies appear and all parameters are as well determined as before.

It is also clear from Fig.~\ref{fig8} (bottom row) that moderate  scatter, up to $\sigma\sim 0.5$ is still consistent with the LF data, within $\sim 1 \sigma$ confidence.  To investigate models with even greater scatter, we also performed two MCMC runs where the scatter was fixed to $0.25$ and $0.50$ dex  (i.e., $\sigma=0.25\ln 10$ and $\sigma=0.50\ln 10$, respectively). The best parameter values for the case of unconstrained $f_{\rm duty}$ for these two runs are shown in the second and third rows in Table~\ref{tab3}. It turns out that the data and model are able to comfortably accommodate $M_{\rm h}-L_{\rm X}$ scatter up to $\sim 0.50$ dex. 

We now investigate the effect of the luminosity dependence of the duty cycle. One possibility of such dependence was suggested by  observations \citep{2008ApJ...681..931B}, as well as by semi-empirical models \citep{2008MNRAS.388.1011M,2013MNRAS.428..421S}. We consider $L_{\rm X}$-dependence in the power law form:
\begin{equation}\label{eq8new}
f_{\rm duty}(L_{\rm X},z)=\left(\frac{L_{\rm X}}{\widetilde{L}_0}\right)^\alpha f_{\rm duty}(z)\,,
\end{equation}
where $f_{\rm duty}(z)$ is given by Eq.~(\ref{eq3}) and $\widetilde{L}_0$ is fixed at $\widetilde{L}_0=10^{43}$ erg/s. As before, we consider the case of unconstrained  $f_{\rm duty}$.

Results of  these calculations are presented in the last two rows of Table~\ref{tab3}, for cases without and with scatter in the $M_h-L_X$ relation. The confidence contours for the model parameters are shown in Fig.~\ref{fig9}, where for comparison we have also plotted our contours from Fig.~\ref{fig8}. Introduction of the $L_X-$dependence of the duty cycle dramatically improves the quality of the fit with the $\chi^2$ reducing by more than $\sim 30-40 \%$. Including the scatter in the $M_h-L_X$ relation does not lead to significant further improvement, although statistically it is justified at the $\sim 1.5-2\sigma$ level. Even though the best fit value of the scatter $\sigma$ is quite high ($\sim 0.6$ dex), its $2\sigma$ contour includes the null value. 

Contrary to what one might have guessed, the best fitting values for $\alpha$ turn out to be positive, i.e., $f_{\rm duty}$ grows with $L_{\rm X}$.\footnote{In fact, we have also performed MCMC runs by explicitly demanding $\alpha$ to be negative along with using several different values for $\widetilde{L}_0$ in Eq.~(\ref{eq8new}). In all of these cases we generally find the quality of fits to deteriorate, the more negative $\alpha$ becomes.}
It is interesting to note that the $L_{\rm X}$-independent part of the duty cycle describing its redshift dependence, $f_{\rm duty}(z)$, (Eq.~(\ref{eq3})), is sufficiently robust and insensitive to the details of the model; it is nearly identical in shape and normalization  to the $f_{\rm duty}(z)$ in our default model in  previous sections. In particular, it is insensitive to the choice of $\widetilde{L}_0$, with changes in $\widetilde{L}_0$ mostly absorbed by corresponding changes in other parameters, mostly by the $\Gamma(z)$ in  Eq.~(\ref{eq1}). Changes in the latter lead to modifying the $M_h-L_X$ relation, such that  the characteristic halo mass $M_*$ is no longer mapped into the characteristic luminosity $L_*$. For this reason, the overall duty cycle $f_{\rm duty}(L_X, z)$ can now take values over unity, for example, $f_{\rm duty}(L_X=10^{46}, z=1)\sim 6$. This should be interpreted as meaning that high mass halos may contain several SMBHs that are in the active state. The robustness and potential implications of this result still need to be explored. 

\begin{figure}
\centering
\includegraphics[width=\plotwd]{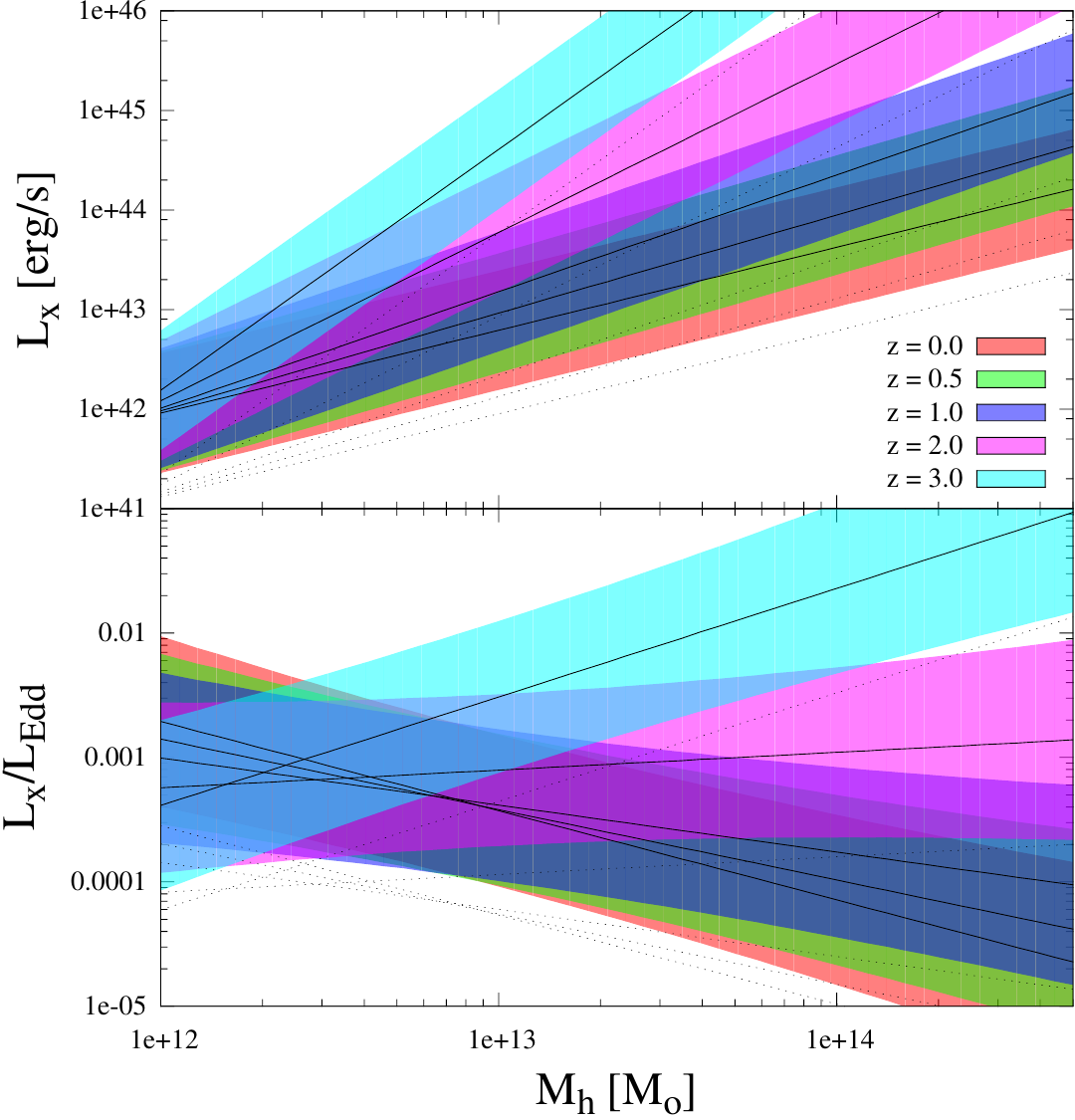}
\caption{Analog of Fig.~\ref{fig6} for the last model in Table~\ref{tab3}, i.e., for the model with $L_{\rm X}$-dependent $f_{\rm duty}$ and with scatter in $M_{\rm h}-L_{\rm X}$ scaling relation. The solid lines surrounded by the shaded areas show distribution medians along with regions between $16\%$ and $84\%$ quantiles, respectively. Dotted lines represent modes of the distributions.}
\label{fig10}
\end{figure}

The  statements of the previous section regarding the low Eddington ratios of X-ray-selected AGN and long mass growth times hold for the extended models, as shown in Fig.~\ref{fig10}. Here the solid lines and the shaded areas show distribution medians and regions between $16\%$ and $84\%$ quantiles, respectively. The strong level of asymmetry in these probability distributions can be seen clearly by looking at the dotted lines, which represent the distribution modes.

\begin{figure}
\centering
\includegraphics[width=\plotwd]{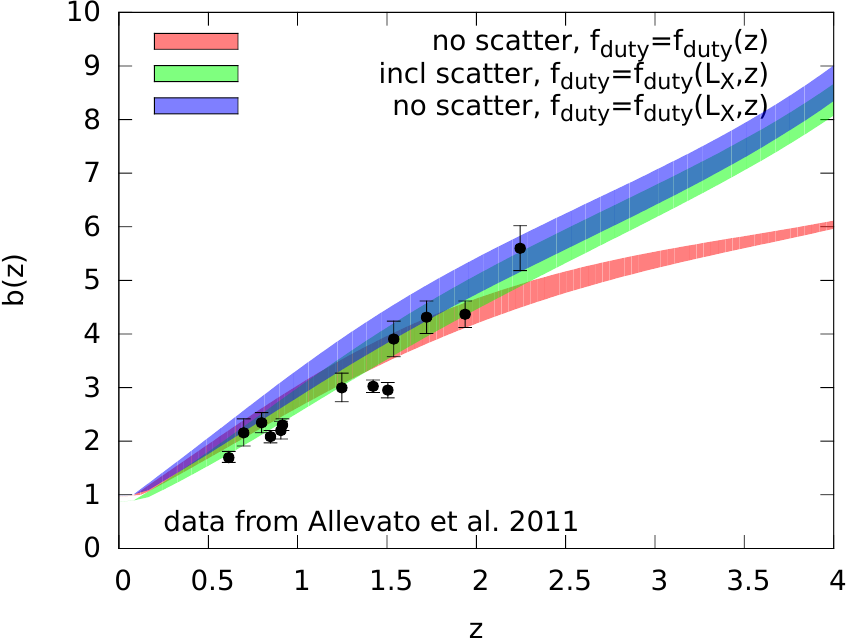}
\caption{Clustering strength as a function of redshift for different models, as shown in legend and discussed in detail in the main text. The filled stripes cover the range of photon indices $\Gamma=1.4-2.0$.}
\label{fig11}
\end{figure}

The shift  of the $M_h-L_X$ mapping toward higher halo masses in the models with luminosity-dependent  $f_{\rm duty}(L_X, z)$, leads to an increase in its predictions for the clustering bias at $z\ga 2$.  This places the new models  in somewhat better agreement with the observational data, as demonstrated in Fig.~\ref{fig11}. The figure compares  predictions of the extended  models with our default model and with results of clustering bias measurements.  The lowest stripe shows the results from Fig.~\ref{fig4}, while the middle and upper ones correspond to the models with luminosity-dependent  $f_{\rm duty}(L_X, z)$, discussed in this section.  While at lower redshifts, $z\la 1.5-2$, the behavior of different models is nearly identical and equally consistent with the data. At $z\ga 1.5$ they start to  diverge, with the extended  models in better agreement with the highest redshift data point. This suggests a possible way to distinguish between different  models, using the clustering bias measurements at redshifts $z\ga 2$. 

Comparing the $\chi^2$ values shown in Tables~\ref{tab2} and \ref{tab3}, we note that  models with a luminosity-dependent duty cycle  give a better fit to the hard band X-ray AGN LF than do the specifically tailored and commonly used mathematical forms like LADE or LDDE. Therefore these models can be used as an easily calculable analytic representation of the LF. To assist in computing the luminosity function,  in Appendixes~\ref{appa}  and \ref{appb} we give a detailed description of the algorithm and provide easy-to-use approximation formulae.

\section{Conclusions}\label{sec4}

Even though dark matter halos are highly nonlinear objects, with the aid of cosmological N-body simulations along with heuristic analytic models, we now have good knowledge about their redshift dependent MF and clustering bias. In CDM type cosmologies, where structures form from the bottom up, one may expect that the buildup of more massive objects has also been accompanied by the formation of more massive SMBH  in their centers. This should also be manifested on average by more luminous AGN. Although there certainly are other key parameters beyond the halo mass, which influence the central SMBH mass, one may expect that $M_{\rm h}$  is one of the main contributors. 

In this paper,  we assumed that  a scaling relation between X-ray AGN luminosity and its host halo mass does exist. Assuming further that the dark matter halo mass function is described by the concordance $\Lambda$CDM model, we can predict X-ray AGN luminosity function. Comparing the latter with the redshift-dependent AGN X-ray luminosity function known from observations we can determine the shape and parameters of the $M_{\rm h}-L_{\rm X}$  scaling relation.

Our main conclusions are the following:
\begin{itemize}
\item We have shown that a simple scaling relation in the form of Eq.~(\ref{eq1}), where $\Gamma(z)$ is a linear or quadratic polynomial, allows one to obtain X-ray AGN LFs consistent with observational data. The best fitting model parameters along with $\chi^2$ values are summarized in Tables~\ref{tab1} and \ref{tab2}.
\item Since the $M_{\rm h}$-$L_{\rm X}$ scaling establishes a link to the halo MF, our model is certainly more predictive than the usual multi-parameter fitting forms tailored  to fit the X-ray LF data. In particular, our model predicts the redshift and luminosity dependence of the X-ray AGN clustering bias, which agrees reasonably with observational data. 
\item The obtained  $M_{\rm h}$-$L_{\rm X}$ scaling relation is in good agreement with the data presented in \citet{2012MNRAS.tmp..101K} and with  semianalytic galaxy-AGN co-evolution models of \citet{2012MNRAS.419.2797F}.
Comparison with semianalytical models suggests that for X-ray AGN, the dominant accretion mode is  the `hot-halo mode', in contrast to the `starburst mode' which is compatible with being a dominant accretion mode for somewhat less clustered but more energetic quasars.
\item The  AGN duty cycle  peaks at the redshift $z\simeq 1$, and its value at the peak is consistent with unity, implying that there is at most one AGN per dark matter halo. 
The  value for the duty cycle close to unity is a consequence of the fact that at the redshift $z\sim 1$, the volume density of X-ray AGN is comparable to the volume density of the dark matter halos:  
$L_{\rm X}\frac{{\rm d}n}{{\rm d}L_{\rm X}}(L_{\rm X}=L_*,z\simeq 1)\simeq M_{\rm h}\frac{{\rm d}n}{{\rm d}M_{\rm h}}(M_{\rm h}=M_*,z\simeq 1)$.
\item We further combined our $M_{\rm h}$-$L_{\rm X}$ relation with the $M_{\rm h}$-$M_{\rm BH}$ relation from \citet{2009ApJ...704.1135B}, along with its redshift evolution taken from \citet{2009MNRAS.394.1109C}, and obtained Eddington ratios $f_{\rm Edd}\equiv L_{\rm X}/L_{\rm Edd}\sim 10^{-4}-10^{-3}$ (2--10 keV band, no bolometric correction applied). This result presents another evidence  that at the redshifts below  $z\la 3$,   SMBHs in X-ray selected AGN typically grow through `hot-halo' accretion mode characterized by low mass accretion rate and a long e-folding time. It is also consistent with the fact that   according to their clustering strengths, X-ray selected AGN populate group-sized dark matter halos and thus are typically embedded in halos of hot  low density gas.
\item We also investigated a class of extended models that include a scatter in the $M_{\rm h}$-$L_{\rm X}$ scaling relation and a luminosity-dependent $f_{\rm duty}$. While the scatter alone is not required by the X-ray LF data, including the luminosity dependence of the duty cycle leads to significant improvement of the fit quality. In the new models, the statements regarding the low Eddington ratios of X-ray-selected AGN and their long SMBH growth times hold. However, they differ in their predictions regarding the behavior of the clustering bias at redshifts $z\ga 2$, with  extended models in better agreement with the most recent bias  measurements at $z\sim 2.2$ by \citet{2011ApJ...736...99A}.
\item In Appendix~\ref{appb}, we summarize the formulae that can be used  to construct an analytical model of the hard-band AGN LF. From the statistical point of view, it provides a better description of the X-ray LF data as specially tailored analytical forms commonly used in the literature.
\end{itemize}

NOTE: during the refereeing process, a new paper by \citet{2013arXiv1305.2200F} appeared, which touches similar topics to the ones discussed in the current study.

\begin{appendix}
\section{Analytic fits for concordance $\Lambda$CDM halo MF and bias parameters}\label{appa}
\begin{figure*}
\centering
\includegraphics[width=\plotwdtwo]{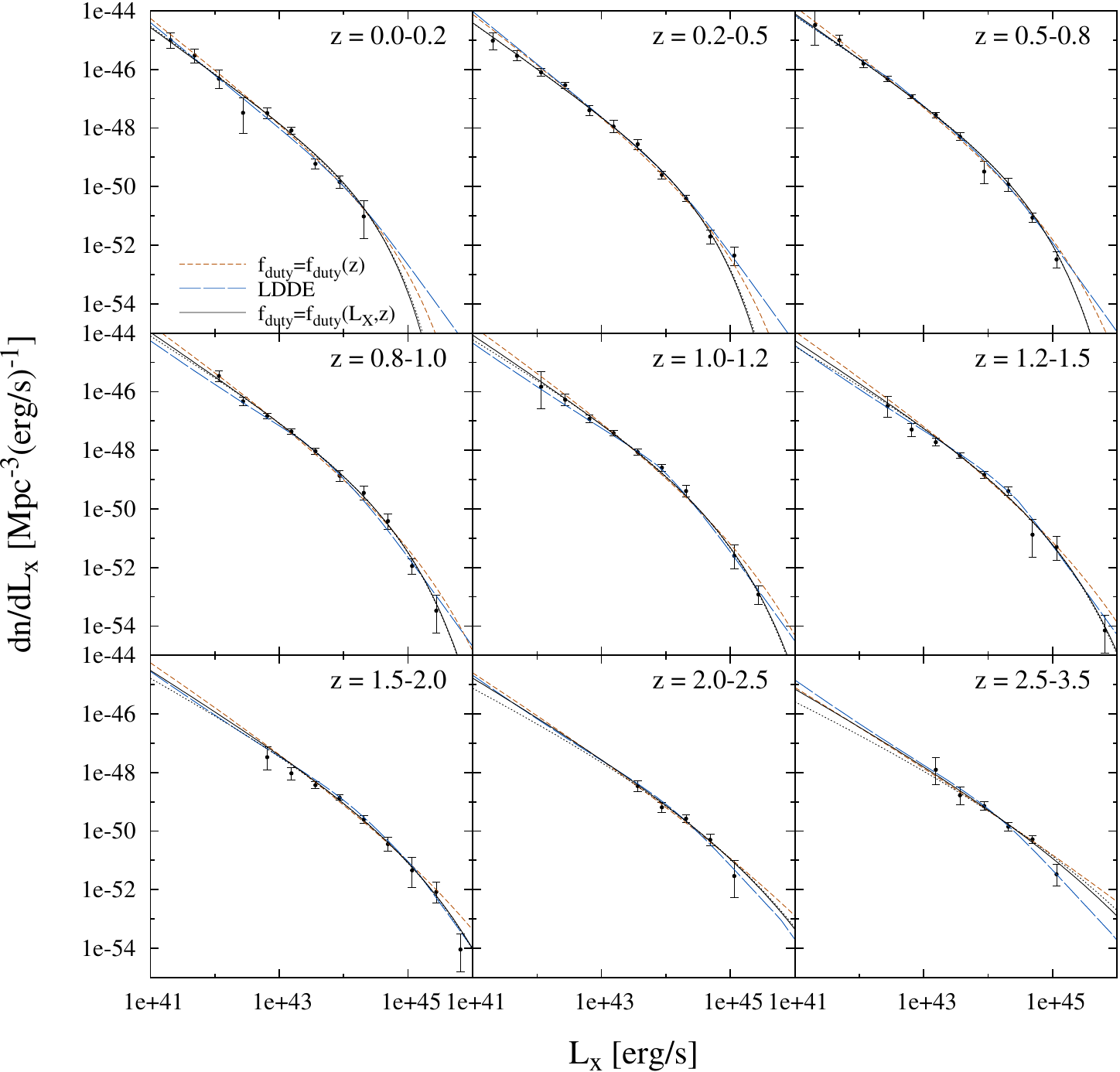}
\caption{Comparison of the LF fit from Appendix~\ref{appb} with the `new model' and LDDE model of Fig.~\ref{fig3}. In addition, with dotted lines we have shown the model with the lowest $\chi^2$ value obtained throughout this paper, i.e., the model with scatter in $M_{\rm h}-L_{\rm X}$ scaling relation along with $L_{\rm X}$-dependent duty cycle (the last entry of Table~\ref{tab3}).}
\label{fig12}
\end{figure*}
To facilitate a fast check of our results for those who do not have numerical routines for the halo MF and clustering bias available, we provide here analytic fits for those quantities, valid for the mass and redshift ranges relevant for this study: $M_{\rm h}\gtrsim 10^{12}\,M_\odot$, $z < 3-4$. 

Halo MF can be approximated as
\begin{eqnarray}
\frac{{\rm d}n}{{\rm d}M_{\rm h}}(M_{\rm h},z) &\simeq& A(z)\left(\frac{M_{\rm h}}{M_*(z)}\right)^{\alpha(z)}\times\nonumber\\
&\times& \exp\left[-\left(\frac{M_{\rm h}}{M_*(z)}\right)^{\beta(z)}\right]\,,\label{eq1app}
\end{eqnarray}
where
\begin{eqnarray*}
A(z) &=& \exp\left(-0.1057\cdot z^2 + 2.677\cdot z - 43.89\right)\,,\\
M_*(z) &=& \exp\left(0.08305\cdot z^2 - 1.709\cdot z + 33.17\right)\,,\\
\alpha(z) &=& 0.007189\cdot z^2 - 0.1692\cdot z - 1.919\,,\\
\beta(z) &=& 0.001109\cdot z^2 - 0.04119\cdot z + 0.7188\,.
\end{eqnarray*}
Similarly, for the halo clustering bias, as a function of halo mass and redshift, one approximately obtains
\[
b(M_{\rm h},z)\simeq c_1(M_{\rm h})\cdot z^2+c_2(M_{\rm h})\cdot z+c_3(M_{\rm h})\,,
\] 
where
\begin{eqnarray*}
c_1(M_{\rm h}) &=& \exp\left[0.01703\ln(M_{\rm h})^2-0.6629\ln(M_{\rm h})+3.566\right] \,,\\
c_2(M_{\rm h}) &=& \exp\left[0.01495\ln(M_{\rm h})^2-0.5196\ln(M_{\rm h})+1.630\right]\,,\\
c_3(M_{\rm h}) &=& \exp\left[0.02866\ln(M_{\rm h})^2-1.534\ln(M_{\rm h})+20.38\right]\,.
\end{eqnarray*}
Under the mass and redshift constraints as given above, those analytic forms provide fits to the \citet{1999MNRAS.308..119S} MF and \citet{2001MNRAS.323....1S} bias parameter with accuracies $\sim 10\%$ and better than $5\%$, respectively. In the above formulae, we have assumed that MF is measured in units of $h^3$Mpc$^{-3}/h^{-1}M_\odot$ and halo mass in $h^{-1}M_\odot$, while in the main part of our paper, we fixed the reduced Hubble parameter to $h=0.7$.

\section{A simple physically motivated fit to the observed hard band X-ray AGN LF}\label{appb}
The hard band X-ray AGN LF can be approximated as
\begin{eqnarray}
\frac{{\rm d}n}{{\rm d}L_{\rm X}}(L_{\rm X},z)&=&f_{\rm duty}(L_{\rm X},z)\Gamma(z)\frac{M_{\rm h}(L_{\rm X},z)}{L_{\rm X}}\times \nonumber \\ &\times& h^3\frac{{\rm d}n}{{\rm d}M_{\rm h}}\left[hM_{\rm h}(L_{\rm X},z),z\right]\,,
\end{eqnarray}
where $f_{\rm duty}(L_{\rm X},z)$, $M_{\rm h}(L_{\rm X},z)$, and $\frac{{\rm d}n}{{\rm d}M_{\rm h}}\left(M_{\rm h},z\right)$ are given by Eqs.~(\ref{eq8new},\ref{eq3}), (\ref{eq1}), and (\ref{eq1app}), respectively. Using the approximation for the MF from Appendix~\ref{appa} leads to slightly different best fitting parameter values compared to the ones shown in the last row of Table~\ref{tab3}: $\{c_i\}_{i=1\ldots 9}=$ {\scriptsize $\{0.3654,0.8243,0.2402,1.002,-0.2997,0.03216,26.97,0.,0.3198\}$}, which results in $\chi^2=47.5$ for $71$ degrees of freedom.

A detailed comparison of the above LF fit with the `new model' and LDDE model of Fig.~\ref{fig3} is shown in Fig.~\ref{fig12}.

\end{appendix}

\acknowledgements{We thank James Aird for providing the tables of hard-band X-ray luminosity function measurements and our referee for suggestions that greatly helped to improve the paper.}
\bibliographystyle{aa}
\bibliography{references}

\end{document}